\documentclass[12pt]{article}

\usepackage{scicite}

\usepackage{setspace}

\usepackage{graphicx}
\usepackage{amsmath}
\usepackage{amssymb}
\usepackage[final]{pdfpages}
\newcounter{firstbib}

\usepackage{times}

\newenvironment{sciabstract}{%
\begin{quote} \bf}
{\end{quote}}

\newcounter{lastnote}
\newenvironment{scilastnote}{%
\setcounter{lastnote}{\value{enumiv}}%
\addtocounter{lastnote}{+1}%
\begin{list}%
{\arabic{lastnote}.}
{\setlength{\leftmargin}{.22in}}
{\setlength{\labelsep}{.5em}}}
{\end{list}}

\title{A radio jet from the optical and x-ray bright stellar tidal disruption flare ASASSN-14li}

\author{\large S. van Velzen$^1$, G.\,E. Anderson$^{2,3}$, N.\,C. Stone$^4$, M. Fraser$^5$,  T. Wevers$^6$, \\
B.\,D. Metzger$^4$, P.\,G.~Jonker$^{7,6}$, A.\,J. van~der Horst$^8$, T.\,D. Staley$^2$,  A.\,J. Mendez$^1$, \\
J.\,C.\,A. Miller-Jones$^3$, S.\,T. Hodgkin$^5$, H.\,C.~Campbell$^5$, and R.\,P. Fender$^2$ \\ 
\footnotesize{$^{1}$}Department of Physics \& Astronomy, The Johns Hopkins University, Baltimore, MD 21218, USA, \texttt{sjoert@jhu.edu}  \\
\footnotesize{$^{2}$}Department of Physics, University of Oxford, Denys Wilkinson Building, Keble Road, Oxford, OX1 3RH, UK \\
\footnotesize{$^{3}$}International Centre for Radio Astronomy Research, Curtin University, GPO Box U1987, Perth, WA 6845, Australia \\
\footnotesize{$^{4}$}Columbia Astrophysics Laboratory, Columbia University, New York, NY, 10027, USA \\
\footnotesize{$^{5}$}Institute of Astronomy, University of Cambridge, Madingley Road, Cambridge, CB3 0HA, UK\\
\footnotesize{$^{6}$}Dept. of Astrophysics, Radboud University Nijmegen, Heyendaalseweg 135, 6525 AJ, Nijmegen, The Netherlands \\
\footnotesize{$^{7}$}SRON, Netherlands Institute for Space Research, Sorbonnelaan 2, 3584 CA, Utrecht, The Netherlands \\
\footnotesize{$^{8}$}Department of Physics, The George Washington University, 725 21st Street NW, Washington, DC 20052, USA \\
}

\date{}

\begin{document} 
\baselineskip24pt
\maketitle

\begin{sciabstract}
The tidal disruption of a star by a supermassive black hole leads to a short-lived thermal flare. 
Despite extensive searches, radio follow-up observations of known thermal stellar tidal disruption flares (TDFs) have not yet produced a conclusive detection. 
We present a detection of variable radio emission from a thermal TDF, which we interpret as originating from a newly launched jet. The multi-wavelength properties of the source present a natural analogy with accretion state changes of stellar mass black holes, which suggests all TDFs could be accompanied by a jet. In the rest frame of the TDF, our radio observations are an order of magnitude more sensitive than nearly all previous upper limits, explaining how these jets, if common, could thus far have escaped detection. 
\end{sciabstract}
\clearpage 

Although radio jets are a ubiquitous and well-studied feature of accreting compact objects, it remains unclear why only a subset of active galactic nuclei (AGN) are radio-loud. A stellar tidal disruption flare (TDF) presents a novel method with which to study jet production in accreting supermassive black holes. These flares occur after perturbations to a star's orbit have brought it to within a few tens of Schwarzschild radii of the central supermassive black hole and the star gets torn apart by the black hole's tidal force. A large amount of gas is suddenly injected close to the black hole event horizon and we therefore anticipate the launch of a relativistic jet as this stellar debris gets accreted \cite{Giannios11,vanVelzen11}. About two dozen TDFs have so far been discovered at soft x-ray, UV, and optical wavelengths, e.g., \cite{Bade96,Gezari09,vanVelzen10}. All of these flares can be described by black body emission, hence their description as {thermal} TDFs. Hard x-ray emission from a relativistic jet launched after a stellar disruption has been observed in three cases \cite{Bloom11,Levan11,Burrows11,Cenko11,Brown15}. These so-called {relativistic} TDFs are readily detected at radio frequencies [the best-studied source, Swift~J1644+57, reached a peak flux of $30$~mJy at 22~GHz \cite{Zauderer11,Berger12}]. Surprisingly, radio observations of thermal TDFs show no signs of equally powerful jets \cite{Bower13,vanVelzen12b}, bringing into question the universality of jet production triggered by large changes in the accretion rate \cite{Fender04}.

On December 2 2014, the All-Sky Automated Survey for Supernovae (ASAS-SN) reported the discovery of \mbox{ASASSN-14li} \cite{Holoien15}, an optical transient with a blue continuum in {\it Swift} UV/Optical Telescope (UVOT) follow-up observations, located in the nucleus of a galaxy at redshift $z=0.021$. These properties prompted this transient to be classified as a potential stellar tidal disruption flare. The source was also detected in {\it Swift} x-ray Telescope (XRT) observations, but only at soft x-ray energies (0.3--1~keV; Fig.~1). We began a radio monitoring campaign with the Arcminute Microkelvin Imager (AMI) at 15.7~GHz 22 days after the first {\it Swift} observation and obtained two observations with the Westerbork Synthesis Radio Telescope (WSRT) at 1.4~GHz [see Supporting Material (SM), table~S1].  The 15.7~GHz light curve shows a monotonic decay (factor 5 decrease in 140 days; Fig.~2), suggesting that we observed the fading of a relativistic outflow that was produced by the impulsive accretion event onto the supermassive black hole.

The host galaxy is detected at 3~mJy in archival radio images at 1.4~GHz. The expected radio flux due to star formation is at most $10^{-3}$~mJy (SM) and we therefore conclude that the pre-flare radio flux is due to an AGN. The only other property of the host that suggests ongoing accretion before the flare in 2014 is a narrow [O~\textsc{iii}] emission line with a luminosity of $L_{[{\rm O\, \textsc{iii}}]}=8 \times 10^{38}\,{\rm erg}\,{\rm s}^{-1}$. This low luminosity implies the AGN was in the radiatively inefficient, jet-dominated mode \cite{Falcke04}.

Based on the detection of an AGN prior to the optical flare, one might infer \mbox{ASASSN-14li} to be a brief period of enhanced activity of the pre-existing accretion disk, but this is inconsistent with nearly all of the observed properties of the flare. First, the very low x-ray black body temperature ($T\approx0.06$~keV), including significant absorption features, is unlike the x-ray properties of any known AGN \cite{Miller15}. Second, the large x-ray flux increase with respect to the archival upper limit (Fig.~1) is seen for less than 0.5\% of sources in a blind all-sky search for x-ray variability \cite{Kanner13}. Third, the factor of 100 increase with respect to the baseline UV flux seen in \mbox{ASASSN-14li} is over an order of magnitude larger than observed in a 3 year monitoring campaign of 663 AGN \cite{Gezari13}. And finally, we found no significant variability in eight years of optical observations of the host galaxy of \mbox{ASASSN-14li} by the Catalina Real-Time Transient Survey. A stellar tidal disruption is therefore the best interpretation for \mbox{ASASSN-14li}.

The x-ray temperature and luminosity of  \mbox{ASASSN-14li} are similar to thermal TDFs discovered with x-ray surveys \cite{Bade96} and can be explained by a newly formed, radiatively efficient accretion disk with an inner radius at a few Schwarzschild radii from the black hole \cite{Ulmer99}.  Its optical/UV properties are also very similar to previous optically-discovered TDFs, which are characterized by a large and constant black body temperature [$T=(2-3)\times 10^4$~K; table~\ref{tab:bb} and fig.~\ref{fig:uggr}].

Thermal TDFs are typically detected at optical/UV or soft x-ray frequencies, but not both (table~\ref{tab:xfol}). This could be explained by the existence of a region at 1000~Schwarzschild radii from the black hole that produces the optical emission via reprocessing of the x-ray photons that originate from the inner accretion disk \cite{Guillochon14}. If the product of the optical depth for x-ray ionization and the covering factor of this region is $\gg 1$, luminous optical emission would be produced while x-rays from the inner disk are obscured. In this model, \mbox{ASASSN-14li} can be explained if the evolution of the reprocessing layer gradually allows the escape of more x-ray photons towards our line of sight. This would explain why the x-ray light curve tracks the theoretical $t^{-5/3}$ fallback rate only after about 100 days into the monitoring campaign (Fig.~2) and why the optical light curve of x-ray dim TDFs show significantly less variability than that of \mbox{ASASSN-14li} [e.g., the x-ray dim TDF PS1-10jh \cite{Gezari12} showed only 0.03~mag root-mean-square variability, compared with 0.2~mag for \mbox{ASASSN-14li}]. An alternative explanation for the initially constant x-ray flux is inefficient circularization of the tidal debris streams \cite{Shiokawa15}, which slows down the formation of the inner accretion disk.

The key property of \mbox{ASASSN-14li} is the detection of variable radio emission. Adopting the standard flat or slightly-inverted radio spectrum \cite{Falcke04} for emission of the original AGN jet, we found that our 15.7~GHz observations were always below the 3~mJy baseline level of this jet. The decaying 15.7~GHz light curve of \mbox{ASASSN-14li} therefore indicates that we have observed the termination of an AGN jet because of an increased accretion rate. If the AGN jet had not been terminated, we would have expected an increase with respect to this baseline level. Without an engine to drive particle acceleration in the AGN jet, the synchrotron luminosity will decrease on a timescale of $\sim 10$~days at 10~GHz. Inverse Compton cooling of the electrons on TDF photons can speed up this decrease by a factor $\sim 10$ (SM). Hence the radio flux of the original AGN jet is unlikely to be a dominant component to our post-flare radio observations.

The combined optical and x-ray luminosity during the first month of observations of \mbox{ASASSN-14li} amounts to a few tens of percent of the Eddington luminosity (adopting a central black hole mass of $10^{6.5}\,M_{\odot}$; SM), compared to $<1\%$ of the Eddington limit prior to the flare. \mbox{ASASSN-14li} shares several properties with the flares produced by stellar mass black holes upon being subjected to similarly large changes in their accretion rates. Accretion onto stellar mass black holes in x-ray binaries (XRBs) occurs in two distinct spectral modes, which are separated by a state change that occurs at a few percent of the Eddington luminosity \cite{Maccarone03}. Radio observations of XRBs consistently show the disappearance of a steady compact jet and the launch of transient ejecta during the change from the non-thermal (hard) state to the thermal (soft) state \cite{Fender04}. In direct analogy with XRBs, the steady jet that existed prior to the infall of material from the tidal disruption has been quenched or suppressed, and the accretion disk spectrum is now dominated by thermal emission. The co-added {\it Swift} XRT data of the TDF shows no evidence for a non-thermal (2--10~keV) component at the level needed to power the pre-flare [O~\textsc{iii}] line luminosity (SM), suggesting that the geometrically thick accretion flow that powered the previous steady jet has collapsed.

The maximum radio luminosity of \mbox{ASASSN-14li} is three orders of magnitude lower than that of Swift~J1644+57 \cite{Zauderer11} and evolves on a much shorter timescale. This immediately implies a large difference in jet power between these two events. The radio light curve of \mbox{ASASSN-14li} can be reproduced by using a model similar to that applied to Swift~J1644+57 \cite{Giannios11}, in which synchrotron shock emission is produced as the transient ejecta decelerate upon interacting with dense gas in the nuclear region surrounding the black hole.  Assuming the ejecta were launched $\sim 20$~days before the first {\it Swift} observation of \mbox{ASASSN-14li} and applying a simple blast wave model yields a total jet energy of $E_j\sim 10^{48}~{\rm erg}$, under the common assumption that 20 and 1$\%$ of the energy dissipated by the shocks is placed into relativistic electrons and magnetic fields, respectively \cite{Giannios11}. This energy is four orders of magnitude lower than the total jet energy of Swift~J1644+57 \cite{Mimica15}. 

By the time of our radio observations, the newly launched jet would have swept up enough matter to slow to mildly relativistic velocities (bulk Lorentz factor $\Gamma_j \approx 2$), causing each lobe to spread laterally in a quasi-spherical manner (similar to a mushroom cloud). The approximately isotropic nature of the radio emission at the time of the observations also implies that a finely tuned viewing angle with respect to the jet axis is not required.  The gas density of $\sim 10^{3}\,{\rm cm}^{-3}$ that is required to decelerate the jet at a characteristic radius of 0.1~pc can be explained by the Bondi accretion flow needed to supply the radiatively inefficient flow that existed before the flare (SM).  The deceleration of the new jet implies it cannot be launched into the funnel cleared by the previous jet, which occurs naturally if the new jet orientation is determined by the angular momentum of the new accretion disk rather than the black hole spin vector. 

Adopting the analogy of tidal disruption events as laboratories for studying accretion physics, we have thus far obtained two well-sampled multi-wavelength experiments with very different outcomes: one yielded a powerful jet (Swift~J1644+57), whereas the second event promptly followed-up at radio frequencies (\mbox{ASASSN-14li}) revealed a much weaker jet.  A common explanation for the wide range of black hole jet efficiency is black hole spin --- powerful jets require higher spin.  This model, however, cannot readily explain the radio light curve of \mbox{ASASSN-14li}, because it would predict that the radio luminosity should increase after the disruption (because the spin remains unchanged, whereas the gas supply is greatly enhanced with respect to the pre-disruption accretion rate).  Besides spin, powerful jets may require a large magnetic flux near the black hole horizon \cite{McKinney13}.  Our observations could suggest that the magnetic flux stored in a pre-existing accretion flow is not tapped efficiently upon the disruption and accretion of a star, contrary to simulation predictions \cite{Kelley14}. 

The majority of radio follow-up observations of thermal TDFs were obtained many years after the peak of the flare.  Our observations are the first to sample the light curve within 30 days of the peak. Combined with the low redshift of \mbox{ASASSN-14li}, this explains why similar jets in previous thermal TDFs have eluded detection (table~S4). In analogy with the consistent production of transient jets during accretion flow state changes of stellar mass black holes, our observations suggest that radio-emitting outflows could be a common feature of all TDFs. Adopting a 5-$\sigma$ detection threshold of 90~$\mu$Jy for a monthly all-sky survey with the Square Kilometer Array at 1.4~GHz \cite{Fender15}, a galaxy density of $5\times 10^{-3}~{\rm Mpc}^{-3}$ and a jet production rate equal to the observed thermal TDF rate [$3\times 10^{-5}~{\rm galaxy}^{-1}\,{\rm year}^{-1}$ \cite{vanVelzen14}] yields a detection rate of $\sim 10^{2}$ thermal TDF jets  per year. Although the non-thermal tidal disruption jets selected by {\it Swift} are much more powerful, they are a smaller subpopulation of all stellar disruptions. In blind radio transients surveys, both types of TDF jets could be detected at roughly similar rates.

\begin{figure}
\includegraphics[trim=0mm 2mm 0mm 5mm, clip, width=1\textwidth]{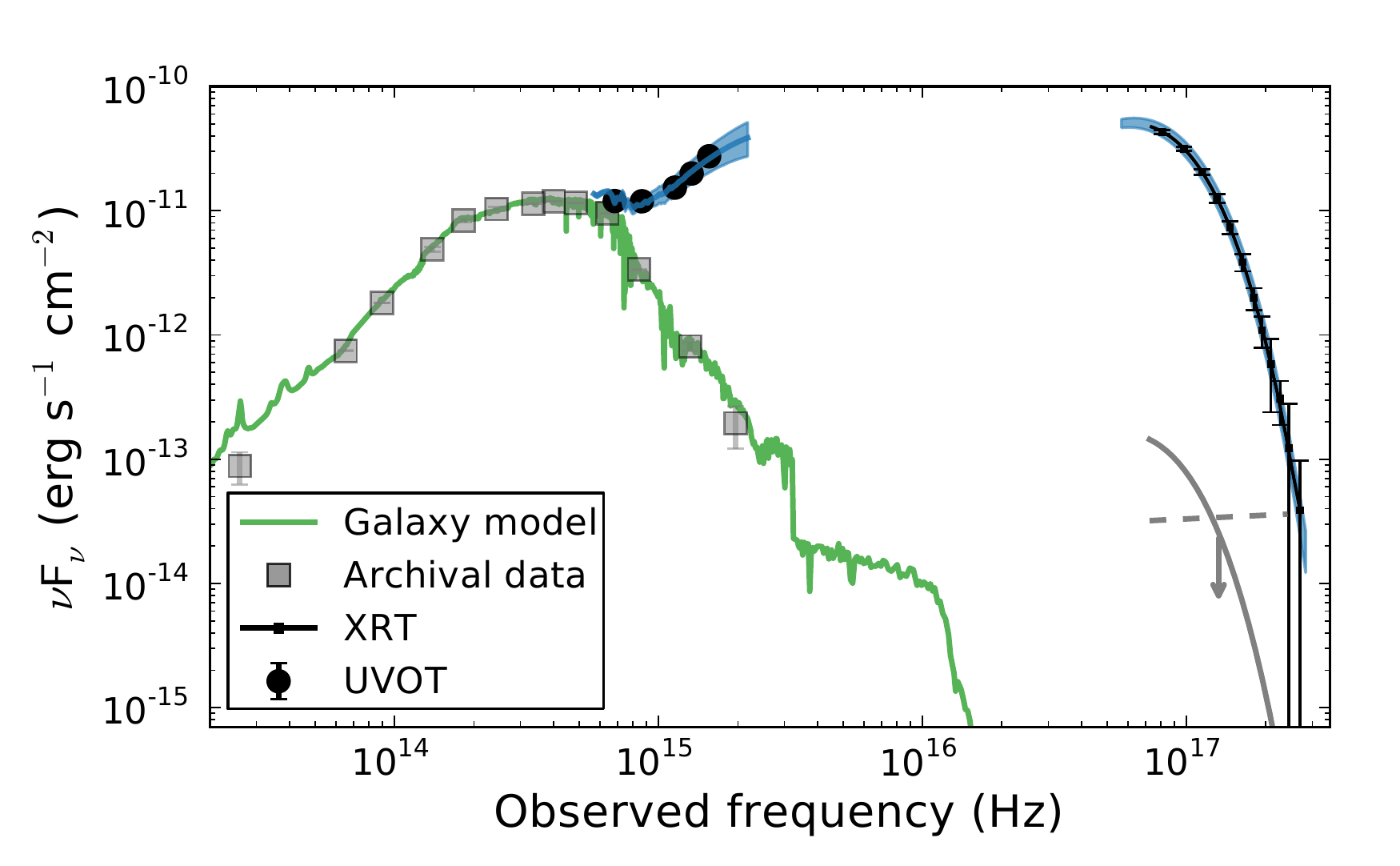}
\caption{{\bf Flare and host spectral energy distribution (SED)}. Shown is the first epoch of the series of {\it Swift} x-ray (unfolded spectrum) and broadband optical/UV observations of \mbox{ASASSN-14li}. These  observations can each be described by a single black body with $T=7.7\times 10^5$~K  and $T=3.5\times 10^4$~K, respectively (blue lines; width reflects uncertainty on the temperature). The SED of the host galaxy based on archival data (gray squares) shows no sign of star formation or an AGN as demonstrated by our best-fit synthetic galaxy spectrum (green line). The pre-flare x-ray limit is shown for both a black body spectrum of similar temperature as the current x-ray spectrum (gray solid line) and a standard power-law AGN spectrum ($\Gamma=1.9$; gray dashed-line). }
\end{figure}
\clearpage

\thispagestyle{empty}
\begin{figure}
\includegraphics[trim=0mm 15mm 0mm 20mm,  width=0.9\textwidth]{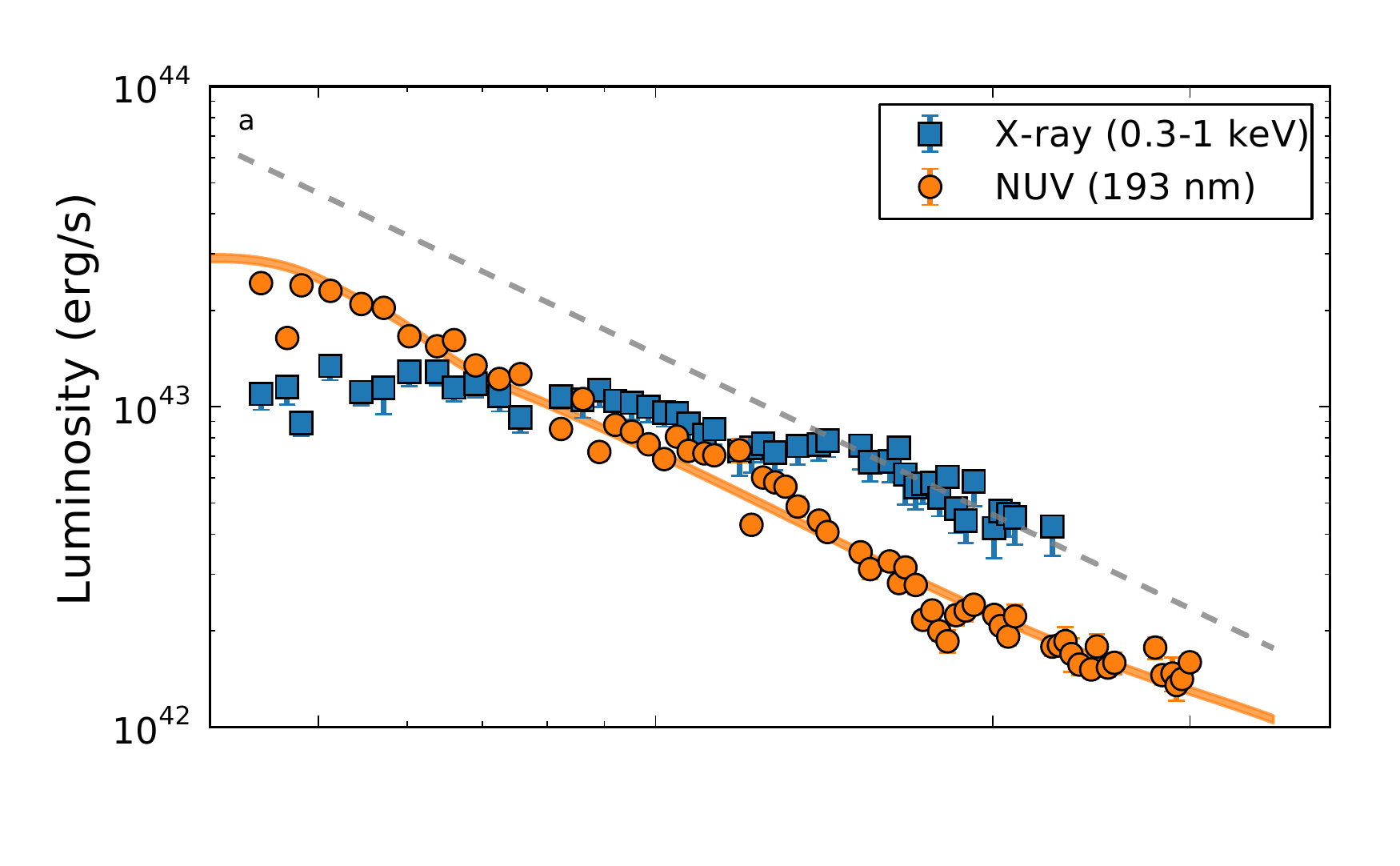}
\includegraphics[trim=0mm 2mm 0mm 5mm, clip, width=0.9\textwidth]{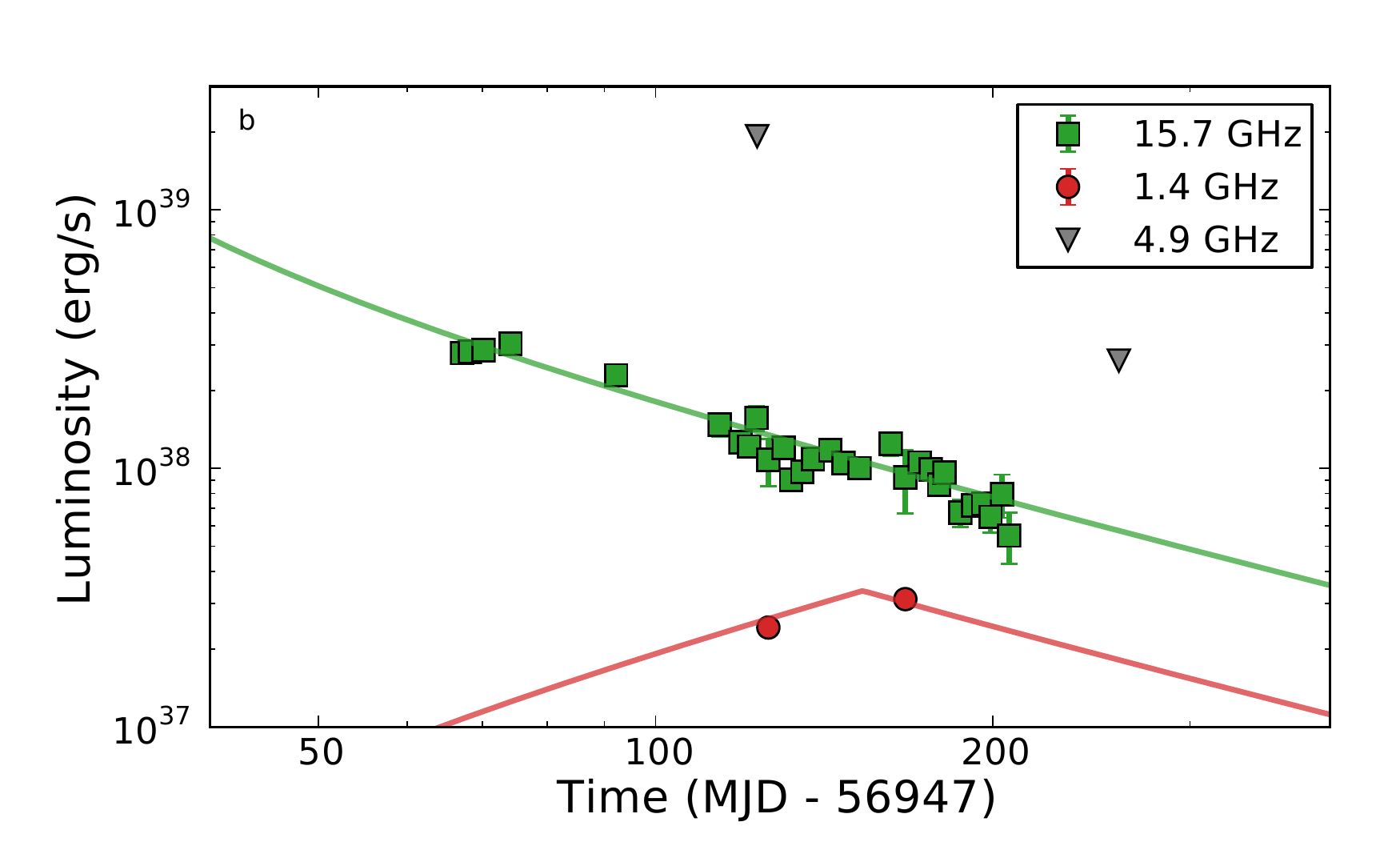}
\caption{{\bf Multi-wavelength light curves of the tidal disruption flare ASASSN-14li.} A. Integrated soft x-ray (0.3--1~keV) luminosity and monochromatic ($\nu L_{\nu}$) near-UV (UVW2-band) luminosity. A spline fit to the observed $g$-band light curve of the known tidal flare PS1-10jh \cite{Gezari12}, corrected for cosmological time dilation and scaled up by 15\%, is shown for reference (solid line). The dashed gray line indicates a $(t-t_{0})^{-5/3}$ power law, the approximate theoretically expected fallback rate of the stellar debris for a disruption at $t_{0}$. The normalization of  the time-axis (in Modified Julian Day, MJD) is chosen to highlight that the (late-time) light curves are consistent with this power-law ($t_{0}=56947\pm 2$~MJD; Fig.~\ref{fig:loglin}). Error bars show the 1-$\sigma$ statistical uncertainty, often smaller than the marker size. B.~Monochromatic radio luminosity at 15.7\,GHz (AMI) and 1.4\,GHz (WSRT) of \mbox{ASASSN-14li} and our jet model (solid lines). The spectral indices during the two epochs of dual radio frequency coverage are $-0.4 \pm 0.1$ and $-0.6 \pm 0.1$ (first and second epoch, respectively). The two most stringent upper limits on the early-time 5~GHz emission of previous thermal TDFs (gray triangles; table~S4) were not sensitive enough to detect transient radio emission similar to \mbox{ASASSN-14li}. }
\end{figure}

\singlespace
\begin{scilastnote}
\item[] {\bf Acknowledgments.} We are grateful to the ASAS-SN team for making their newly identified optical transient public. We thank J. Krolik for useful discussions. We thank the staff of the Mullard Radio Astronomy Observatory for their invaluable assistance in the operation of AMI. We thank the WSRT director for granting the observations in Director's discretionary Time, and the WSRT staff for obtaining these observations. The WSRT is operated by ASTRON (Netherlands Institute for Radio Astronomy) with support from the Netherlands foundation for Scientific Research. SVV is supported by NASA through a Hubble Fellowship (HST-HF2-51350.001). GEA, TDS and RPF acknowledge support from the European Research Council via Advanced Investigator Grant 267697. GEA also acknowledges the support of the International Centre for Radio Astronomy Research (ICRAR), a Joint Venture of Curtin University and The University of Western Australia, funded by the Western Australian State government. MF and HC acknowledge support from the European Union FP7 programme through ERC grant 320360. BDM and NS acknowledge support from NASA grant NNX14AQ68G, NSF grant AST-1410950, and the Alfred P. Sloan Foundation. NS is also supported by NASA through an Einstein Fellowship. JCAMJ is supported by an Australian Research Council Future Fellowship (FT140101082).
\item[]  The data presented here can be found in the SM; raw optical/UV/X-ray observations are available in the NASA/{Swift} archive \\
(http://heasarc.nasa.gov/docs/swift/archive, Target Name: \verb BRUTUS6984_2 ); raw radio observations (WSRT and AMI) are maintained by the observatories and available upon request.
\item[] {\bf Supporting Materials} \\
www.sciencemag.org \\ 
Figs. S1 and S2\\ 
Tables T1-T5\\
References (31-104) \\
\end{scilastnote}
\normalsize

%
%
%
%

\includepdf[pages=-]{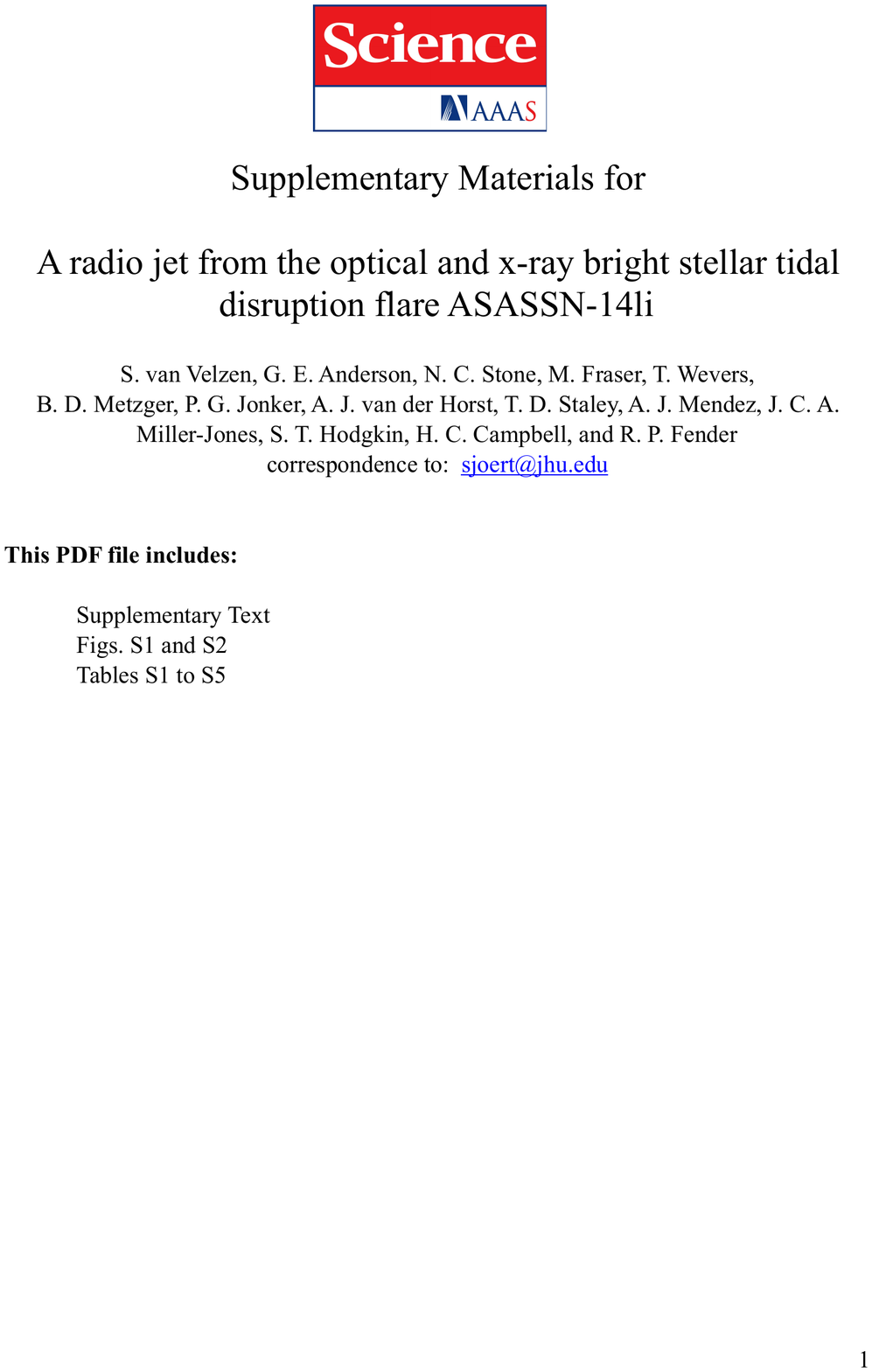} 

\setcounter{page}{1}
\renewcommand{\thefigure}{S\arabic{figure}}
\renewcommand{\thetable}{S\arabic{table}}
\setcounter{figure}{0}

\singlespace
%
%
\section{Radio Observations}
%
%

\mbox{ASASSN-14li} is coincident with the nucleus of the galaxy PGC~043234 (12:48:15.2 +17:46:26, J2000), which is a known faint radio emitter with an archival 1.4 GHz flux from the NRAO VLA Sky Survey \cite{Condon98} or NVSS (NVSS 124815+174629, flux $S_{1.4} = 3.2 \pm 0.4$ mJy, mean time of observations 50105.2 MJD corresponding to $19$ years before outburst) and the Faint Images of the Radio Sky at Twenty centimeters \cite{White97,helfand15} or FIRST survey (FIRST J124815.2+174626, flux $S_{1.4} = 2.67 \pm 0.15$~mJy, mean time of observations 51507.6 MJD corresponding to $15$ years before outburst). The archival data show no sign of significant radio variability between these two epochs \cite{Ofek11b}. 

\mbox{ASASSN-14li} was first observed with the Arcminute Microkelvin Imager Large Array \cite{zwart08} or AMI-LA on 2014-12-23.08 (57014.1 MJD), which detected a radio source coincident with PGC 043234, and consequently \mbox{ASASSN-14li}, yielding a peak flux of $1.93 \pm 0.11$ mJy at 15.7~GHz. \mbox{ASASSN-14li} was monitored on nearly a weekly basis with the majority of observations lasting approximately 4 hours. 

The AMI-LA consists of eight 12.8m dishes that have baseline lengths of \mbox{18-110m}, with an effective frequency range of 13.9-17.5 GHz when using the 0.72 GHz bandwidth channels 3-7 (channels 1, 2, and 8 are disregarded due to their current susceptibility to interference). The AMI data reduction was performed using the fully-automated calibration and imaging pipeline \textsc{AMIsurvey} \cite{Staley15,staley15a}. 

For the calibration stage, \textsc{AMIsurvey} makes use of the AMI specific software package \textsc{reduce}, which is designed to take the raw AMI-LA data and automatically flag for interference, shadowing, and hardware errors, conduct phase and amplitude calibrations, and Fourier transforms the data into \textit{uv}-FITS format \cite{perrott13}. Short interleaved observations of J1255+1817 were used for phase calibration. For the imaging stage, \textsc{AMIsurvey} then calls out to an automated imaging algorithm, \textsc{chimenea}, which is built upon the Common Astronomy Software Applications package or CASA \cite{jaeger08} and is specifically designed to deal with multi-epoch radio observations of transients \cite{staley15b}. The integrated fluxes were then measured from the resulting images using the Low Frequency Array (LOFAR) Transients Pipeline or TraP, which is a pipeline designed by the LOFAR Transients Key Science Project (TKP; http://www.transientskp.org) to identify transients in radio data [see \cite{swinbank15}]. All flux errors output by the TraP are based on Gaussian statistics \cite{Condon98}. 

After observing a factor of two decrease in the AMI-LA light curve of \mbox{ASASSN-14li} during January and February of 2014 we decided it was prudent to obtain radio observations at 1.4 GHz to quantify the radio variability of the core of PGC 043234, which can be compared to the baseline flux values from NVSS and FIRST that predate \mbox{ASASSN-14li}. We applied for director's discretionary time on the Westerbork Synthesis Radio telescope (WSRT, comprising 14 dishes of 25m diameter) and obtained two 12 hour observations at 1.4 GHz on 2015-02-19.85 (57072.9 MJD) and 2015-04-01.74 (57113.7 MJD); 9 of the 14 dishes were operational. The flux and phase calibrations were performed using 3C147 for both observations. The data reduction and analysis were performed with the Multi-channel Image Reconstruction, Image Analysis, and Display \cite{sault95} or MIRIAD software package using standard techniques, with the integrated fluxes being measured using the \textsc{imfit} task. The flux errors correspond to the root-mean-square (RMS) of the intensity as measured with the \textsc{imsad} task. We also remeasured the flux of the radio source in the FIRST data using the same MIRIAD tools. By analyzing all the 1.4 GHz images with the same software we can eliminate any difference that may be caused by the software package rather than intrinsic variability. Our earliest WSRT observation clearly shows a drop in the flux level when compared to the archival value from FIRST; the second WSRT observation shows the flux to have increased by 40\%. AMI-LA observations were also scheduled to coincide with these two WSRT observations to obtain simultaneous spectral information. Both of these AMI-LA observations were rain affected. Though a rain correction has been applied to these data there still may be some residual uncompensated flux attenuation. The large error bars on these flux values reflect the uncertainty in these measurements. The spectral indices during the two WSRT observations were  $-0.4 \pm 0.1$ (57072 MJD) and  $-0.6 \pm 0.1$ (57113 MJD). The details and results of our AMI and WSRT observations, and the new flux measurement of the archival FIRST observation, are given in table~\ref{tab:radio}.

\begin{table}
\small
\caption{ {\bf Radio observations of ASASSN-14li.} }
\label{tab:radio}
 \begin{center}
 \begin{tabular}{cccccc}
 \hline
Obs. start & Int. time & Facility & Frequency & Peak flux density \\
 (MJD) & (hrs) & & (GHz) & (mJy) \\
\hline \hline
  51505.66 &  47.0$^{\mathrm{*}}$  &  FIRST  &	1.4  &      $2.55  \pm   0.14$\\
  57014.08 &  4.0    &  AMI-LA &	15.7 &     $1.93  \pm   0.11$       \\ 
  57015.17 &  4.0    &  AMI-LA &	15.7 &     $1.95  \pm   0.13$        \\
  57017.10 &  4.0    &  AMI-LA &	15.7 &     $1.98  \pm   0.11$        \\
  57021.11 &  4.0    &  AMI-LA &	15.7 &     $2.10  \pm   0.12$        \\
  57039.09 &  4.0    &  AMI-LA &	15.7 &     $1.58  \pm   0.09$        \\
  57061.05 &  1.0    &  AMI-LA &	15.7 &     $1.02  \pm   0.10 $        \\
  57066.00 &  2.0    &  AMI-LA &	15.7 &     $0.87  \pm   0.07$        \\
  57068.09 &  4.0    &  AMI-LA &	15.7 &     $0.84  \pm   0.06$        \\
  57069.97 &  2.0    &  AMI-LA &	15.7 &     $1.08  \pm   0.12$        \\
  57072.85 &  12.0  &  WSRT  &	1.4  &      $1.91  \pm   0.06$\\
  57072.98 &  4.0    &  AMI-LA &	15.7 &     $0.74  \pm   0.15$      \\  
  57077.06 &  3.0    &  AMI-LA &	15.7 &     $0.83  \pm   0.06$        \\
  57079.06 &  3.0    &  AMI-LA &	15.7 &     $0.62  \pm   0.06$        \\
  57082.11 &  4.0    &  AMI-LA &	15.7 &     $0.67  \pm   0.05$        \\
  57085.09 &  4.0    &  AMI-LA &	15.7 &     $0.75  \pm   0.05$        \\
  57090.08 &  4.0    &  AMI-LA &	15.7 &     $0.81  \pm   0.08$        \\
  57094.06 &  4.0    &  AMI-LA &	15.7 &     $0.72  \pm   0.06$        \\
  57098.95 &  4.0    &  AMI-LA &	15.7 &     $0.69  \pm   0.06$        \\
  57109.04 &  4.0    &  AMI-LA &	15.7 &     $0.86  \pm   0.09$        \\
  57113.74 &  12.0  &  WSRT   &	1.4   &     $2.46  \pm   0.05$\\
  57113.98 &  4.0    &  AMI-LA &	15.7 &     $0.64  \pm   0.18$      \\  
  57118.98 &  4.0    &  AMI-LA &	15.7 &     $0.73  \pm   0.07$        \\
  57122.94 &  4.0    &  AMI-LA &	15.7 &     $0.68  \pm   0.06$        \\
  57125.98 &  4.0    &  AMI-LA &	15.7 &     $0.60   \pm   0.05$        \\
  57127.97 &  4.0    &  AMI-LA &	15.7 &     $0.66  \pm   0.05$        \\
  57133.96 &  4.0    &  AMI-LA &	15.7 &     $0.47  \pm   0.06$        \\
  57138.94 &  4.0    &  AMI-LA &	15.7 &     $0.49  \pm   0.05$        \\
  57142.95 &  4.0    &  AMI-LA &	15.7 &     $0.50   \pm   0.05$        \\
  57145.92 &  4.0    &  AMI-LA &	15.7 &     $0.45  \pm   0.06$        \\
  57150.75 &  4.0    &  AMI-LA &	15.7 &     $0.55  \pm   0.10 $        \\
  57153.73 &  4.0    &  AMI-LA &	15.7 &     $0.38  \pm   0.09$        \\
\hline
\end{tabular}
\end{center}
$^{\mathrm{*}}$ Timespan of the observations.\\
\end{table}

%
%
\section{Host galaxy and AGN}\label{sec:host}
%
%

In this section we present the archival data that allow us to infer the properties of the host galaxy of \mbox{ASASSN-14li} and the pre-existing AGN, prior to the stellar tidal disruption flare (TDF).

The host galaxy of ASASSN-14li has been observed over 270 times on a 8 year baseline by the optical Catalina Real-Time Transient Survey \cite{Drake09}. Our analysis of the cataloged data (DR2) reveals no large outburst prior to the TDF. We computed the inverse-variance weighted mean magnitude of each year, finding a root-mean-square (rms) variability over the 8 year baseline of 0.03~mag.

The broad-band SED of the host galaxy, shown in Fig.~1 of the main text,  includes data from:
\begin{itemize}
\item The Wide Field Infrared Survey Explorer \cite{Wright10}. We used unWISE forced photometry \cite{Lang14} for the W1, W2 and W3 bands. 
\item Two Micron All Sky Survey \cite{Skrutskie06}, using a 6" aperture to extract the total galaxy flux from the J- H- and K-band images. 
\item SDSS $u$, $g$, $r$, $i$, $z$ model magnitudes \cite{stoughton02}.
\item Galaxy Evolution Explorer \cite{Martin05},  using a 6" aperture to extract the flux from the NUV and FUV images.
\item ROSAT All Sky Survey \cite{Voges99}. We used both a power-law with index $\Gamma=1.9$ and a black body spectrum with $T=0.06$~keV to convert the 90\% CL upper limit on the count rate (0.005~counts~s$^{-1}$, based on the exposure map at the location of the source) to a limit on the flux, using a hydrogen column depth of $4\times 10^{20}~{\rm cm}^{-2}$ \cite{Miller15}.
\end{itemize}

All magnitudes are corrected for Galactic dust extinction using $E(B-V)=0.024$ \cite{Schlegel98} and a Fitzpatrick extinction curve \cite{Fitzpatrick99,Yuan13}. To estimate the stellar mass and the star formation rate (SFR), we fit the host galaxy SED with a population synthesis model using \textsc{isedfit} \cite{Moustakas13}. \textsc{isedfit} is a Bayesian fitting code that compares a large Monte Carlo grid of SED models, which span a wide range of stellar population parameters (e.g., age, metallicity, and star formation history), to the observed photometry to estimate the host galaxy properties. Stochastic bursts of star formation of varying amplitude, duration, and onset time are superimposed, allowing for a wide range of possible star formation histories \cite{Kauffmann03b,Salim07}. \textsc{isedfit} marginalizes the full posterior probability distribution of stellar masses and SFRs over all other parameters and thus encapsulates both the uncertainties in the observations and the model parameter degeneracies. The reduced $\chi^2$ of the best-fit model is $0.85$. We take the median stellar mass and SFR from the full probability distribution functions as the best estimate of the stellar mass and SFR and find $(2.9 \pm 0.01)\times 10^9\,M_\odot$ and $10^{-2.8\pm0.7}\, M_\odot \,{\rm yr}^{-1}$, respectively. 

The velocity dispersion of the stellar absorption lines in the SDSS spectrum is $\sigma=61\pm 4\,{\rm km}\,{\rm s}^{-1}$ \cite{Abazajian09,Bolton12}, from which we estimate a black hole mass (BH) of $M_{\rm BH} =10^{6.2\pm 0.3}\,M_\odot$ \cite{Kormendy13}, with the uncertainty in the exponent reflecting the scatter in the $M$--$\sigma$ relation. The velocity dispersion is below the SDSS spectral resolution for a single line, but is reliable because multiple absorption lines are detected at high signal-to-noise ratio (SNR). The mass of the central BH of the host of \mbox{ASASSN-14li} can also be estimated from the total stellar mass. We use the bulge-to-total-ratio of \cite{Lackner12}, $0.82\pm 0.03$ in the $i$-band, to find a BH mass of $M_{\rm BH} = 10^{6.8\pm 0.3}\,M_\odot$ \cite{Kormendy13}. We adopt $10^{6.5}\,M_\odot$ as our fiducial mass estimate.

The SFR can be used to estimate the radio luminosity \cite{Condon02}, 
$$
 {\rm SFR}(M_\odot \,{\rm yr}^{-1}) = 7.8 \times 10^{-37} L_{\rm 1.4\,GHz}\,({\rm erg}\,{\rm s}^{-1}) \quad.
 $$
From the SFR estimate based on the stellar population synthesis we find a predicted 1.4~GHz radio flux of $S_\nu=10^{-2.9\pm0.7}$~mJy. This low radio flux is clearly inconsistent with the 3~mJy radio detections obtained years before the outburst (from NVSS and FIRST, see previous section). 

Besides the pre-flare radio detection, a second indicator that the host of \mbox{ASASSN-14li} harbors an AGN is the detection of [O~{\sc iii}] line emission in the SDSS spectrum; we measured a luminosity of $(8.1\pm0.7)\times10^{38}\,{\rm erg}\,{\rm s}^{-1}$. The detection of [O~{\sc iii}] without any signature of H$\beta$ emission implies that high-energy photons of non-stellar origin are responsible for the line emission \cite{baldwin81}. The luminosity of this continuum can be estimated from the correlation between hard x-ray (2--10~keV) and [O~{\sc iii}] luminosity as observed for type-I AGN \cite{Heckman05}, yielding  $L_X = 10^{41.0\pm 0.5}\,{\rm erg}\,{\rm s}^{-1}$. This pre-flare x-ray luminosity is consistent with the upper limit on the flux in this energy range, $L_X<10^{40.7}\,{\rm erg}\,{\rm s}^{-1}$ based on the ROSAT non-detection. 

Applying a bolometric correction \cite{Hopkins07} to the ROSAT upper limit on the x-ray luminosity, we find an upper limit on the disk luminosity of a radiatively efficient accretion disk (i.e.,  broad-line quasar, or type-1 AGN) of $L_{\rm bol}<10^{41.7\pm 0.3}\,{\rm erg}\,{\rm s}^{-1}$. When we convert the [O~{\sc iii}] line luminosity to a bolometric disk luminosity, using the correlation observed for type-I AGN \cite{heckman04}, we find $L_{\rm bol}=10^{42.4\pm 0.4}\,{\rm erg}\,{\rm s}^{-1}$. While the conversion between [O~{\sc iii}] luminosity and the non-thermal continuum flux that is ionizing this line can be expected to remain roughly valid at low accretion rates, the conversion to bolometric luminosity is a strict upper limit since the AGN is not likely to be radiatively efficient at such a low Eddington ratio. The bolometric disk luminosity estimated from the [O~{\sc iii}] data is inconsistent with the limit on the disk luminosity from the broad-band host galaxy SED, which we estimate to be $L_{\rm bol}<10^{41.5\pm 0.3}$ (based on the lack of blue-bump emission at $<10\%$ of the host galaxy light in the $u$-band). We are thus driven to conclude that the AGN that existed before the stellar tidal disruption was radiatively inefficient and accreting at a rate less than 1\% of the Eddington limit.

%
%
\section{Optical and UV observations}
%
%

The {\it Swift} satellite observed the field of \mbox{ASASSN-14li} starting 2014 November 30, 8 days after the discovery of the transient by ASASSN \cite{Holoien15,Jose14}. Here we report on the properties derived from the {{\it Swift} UV Optical Telescope (UVOT) data obtained between 2014 November 30 and 2015 August 12.

The UVOT observations were reduced following the standard procedures and software (\textsc{uvotsource}). We used a 4" aperture and the B, U, UVW1, UVM2, and UVW2 filters, spanning 1928--4392~\AA   \cite{Poole08}. We present the results of our fit for the (rest-frame) black body temperature and radius of the flare in table~\ref{tab:bb}.  Since the host galaxy flux in the UVOT filters is unknown, we used our galaxy model (see previous section) to compute the flux from the host plus a black body and convolved this spectrum with the UVOT filters. We adopted a 5\% uncertainty on our estimate of the host flux. 

The luminosity integrated over the UVOT bands is $(2.6\pm 0.1) \times 10^{43}\,{\rm erg}\,{\rm s}^{-1}$ at the first epoch of observations; the peak black body luminosity is $(5.4 \pm 2.4) \times 10^{43}\,{\rm erg}\,{\rm s}^{-1}$. 

If the light curve of ASASSN-14li is similar to the well-sampled TDF PS1-10jh, the peak occurred about 8 days before the first UVOT observation (Fig.~\ref{fig:loglin}). This inference is consistent with the Liverpool Telescope $g$-band data that indicates the flare was caught at or before its peak \cite{Holoien15}. If we fit the UVW2 light curve of \mbox{ASASSN-14li} with a power-law, $F  \propto (t-t_0)^p$, we find $t_0=-39 \pm 9$~days since the first UVOT observation and $p=-1.71 \pm 0.14$, or $p=(-5.1\pm0.4) / {3}$. The entire NUV light curve is well fit by this power law; the most recent epochs of observation demonstrate that it is not described by an exponential decay. This is in contrast to the initial conclusion of \cite{Holoien15}, who analyzed a subset of this light curve, but is consistent with the majority of optically-selected TDFs.

\begin{figure}[t]
\centering
\includegraphics[trim=0mm 1mm 0mm 1mm,  width=1\textwidth]{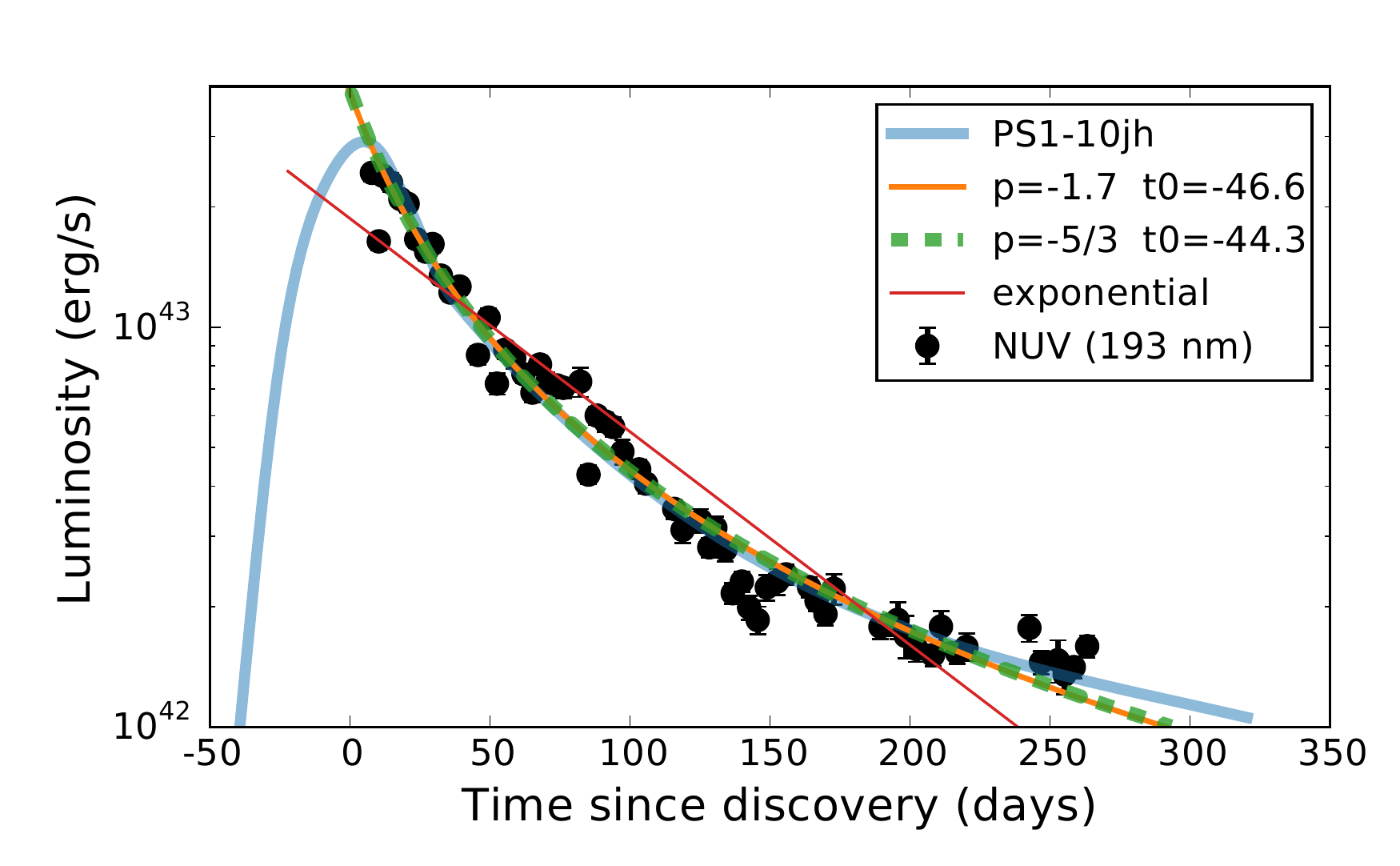}
\caption{{\bf Light curve of ASASSN-14li on a log-linear scale}. We compare the light curve to TDF PS1-10jh (solid blue line) and several simple models. On this log-linear scale an exponential decay ($L\propto e^{-t}$) appears as a straight line, which  does not describe the observations. Instead, a single power law with index $p=-5/3$ provides a decent fit (green dashed lines), the rms scatter to this fit is 0.16~mag. If we leave the power-law index  as a free parameter (solid orange line), we find that a relatively narrow range of indices is consistent with the data: $p=-1.7\pm0.1$. } \label{fig:loglin}
\end{figure}

%
%
\section{X-ray observations}
%
%

Here we report on results obtained using the {\it Swift} X-ray Telescope (XRT) data between 2014 November 30 and 2015 August 12.  The {\sl ftools} software package tool {\sl xselect} has been used to extract source and background photons from regions centered on the best-fit source position and a source free region on the CCD, respectively. All observations were obtained in photon--counting mode where two-dimensional information is available. Circular extraction regions with radii of 59 and 71 arcseconds were used for the source and background regions, respectively. Events were selected if their event-grade was 0--12. We used the response file \verb swxpc0to12s6_20130101v014.rmf   during our spectral analysis while ancillary response files were created for each observation separately using the {\sl ftools} command {\sc xrtmkarf}. We found that the typical XRT King profile \cite{Kingprofile} provides a good description for the observed PSF, hence pile-up is unlikely to be a problem for our observations. 

The extracted spectra were analyzed using the {\sc XSPEC} package \cite{Arnaud96} version 12.8.2. For the fit statistic, we use Cash statistics \cite{Cash79} modified to account for the subtraction of
background counts, the so called W-statistics \cite{Wstat}. We have determined spectral parameters of a black body model restricting the fits to photon energies in the range from 0.3--7~keV and adopted a hydrogen column depth of $4\times 10^{20}~{\rm cm}^{-2}$ \cite{Miller15}.

The x-ray light curve shows only very minor evolution in luminosity and spectral shape (table~\ref{tab:lc}). The median integrated x-ray luminosity from 0.3--7~keV is $7 \times 10^{42}\,{\rm erg}\,{\rm s}^{-1}$ and the median  black body luminosity is $6 \times 10^{43}\,{\rm erg}\,{\rm s}^{-1}$.  The sum of the integration time of the XRT exposures is 144~ks. 
For energies larger than 2~keV, only 16~photons are detected, allowing us to set a 90\% CL upper limit on the count rate of $1.1\times 10^{-4}\,{\rm counts}\,{\rm s}^{-1}$. For a photon index of $\Gamma=2$, this upper limit on the count rate implies a maximum hard x-ray (2--10~keV) luminosity of $L_X< 10^{39.7}\,{\rm erg} \,{\rm s}^{-1}$. This limit is not consistent with the expected hard x-ray luminosity that is needed to power the pre-flare [O~{\sc iii}] line luminosity ($L_X<10^{41.0 \pm 0.5}\,{\rm erg} \,{\rm s}^{-1}$; see section~\ref{sec:host}), suggesting the source of the hard x-ray photons has disappeared, or at least become significantly fainter, after the optical flare.

%
%
\section{Comparison to other TDFs}\label{sec:comparison}
%
%

\begin{table}
\caption{ {\bf Temperature and luminosity evolution of ASASSN-14li and other TDFs.}}\label{tab:lc}
\begin{center}
\begin{tabular}{l l c c c}
\hline
TDF name  & Instrument	& $d \ln T/dt$ & $d \ln L /dt$ & Band \\ 
          	                     & & ($\times 10^{-3}$~d$^{-1}$) & ($\times 10^{-2}$~d$^{-1}$) \\
\hline \hline          	                     
\mbox{ASASSN-14li}  &XRT  &  $-0.9 \pm 0.1$ & $-0.59 \pm 0.04$ & 0.3--1~keV \\
\mbox{ASASSN-14li}  &UVOT & $-0.3 \pm 0.4 $ & $-1.91 \pm 0.02$ & $U$ \\[8pt]

PS1-11af\cite{Chornock14} & PS1 & $\phantom{-}0 \pm 2 $ & $-1.8\pm 0.2$ & $g_{\rm PS1}$ \\
SDSS-TDE1\cite{vanVelzen10} & SDSS & $-2 \pm 4$ & $-1.7\pm0.1$ & $g$ \\
SDSS-TDE2\cite{vanVelzen10}  & SDSS & $-3 \pm 2 $ & $-0.8\pm0.1$ & $g$ \\ 
\hline
\end{tabular}
\end{center}
\end{table}

\begin{figure}[t]
\centering
\includegraphics[trim=5mm 0mm 0mm 10mm, clip, width=1\textwidth]{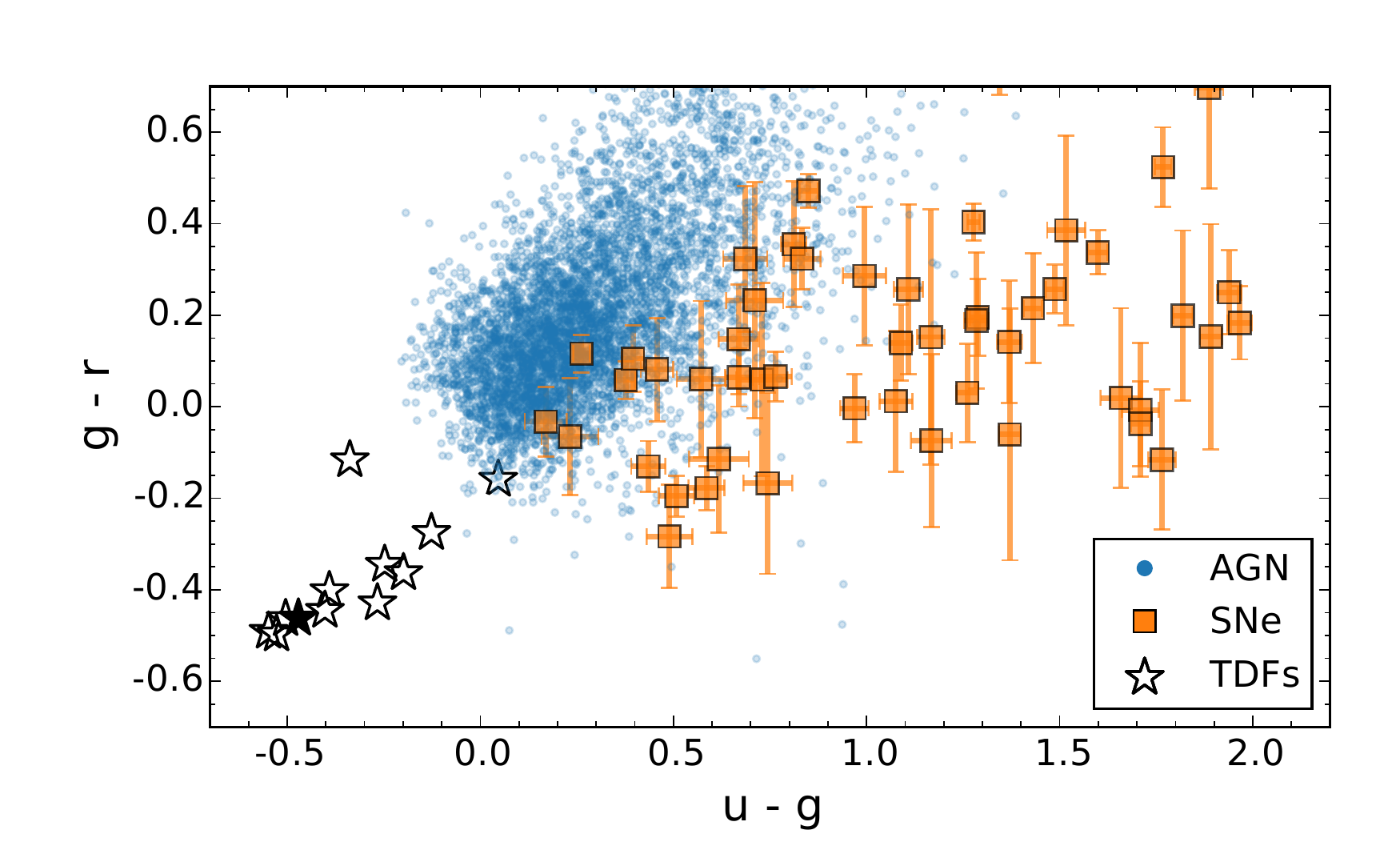}
\caption{{\bf ASASSN-14li in the locus of TDFs}. This color-color diagram shows 12 known tidal disruption flares candidates that have been found in UV/optical imaging surveys, \mbox{ASASSN-14li} is shown by the filled star symbol. The supernovae are from an unbiased sample of transients within 1" of their host galaxy \cite{vanVelzen10} and the AGN shown here are spectroscopically-confirmed quasars from SDSS \cite{schneider10,Ross12} with $z<0.4$. } \label{fig:uggr}
\end{figure}

\begin{table}
\small
\begin{center}
\caption{{\bf x-ray follow-up  observations of TDFs discovered in UV/optical surveys}}\label{tab:xfol}
\begin{tabular}{l c c c l}
\hline
TDF name & $\Delta T$ & $L_{X}$ & Energy & Notes \\
	& (days)  & (erg~s$^{-1}$)& (keV)  &  \\
\hline \hline
\mbox{ASASSN-14li} 	& 10--100 & \phantom{$<$} $1 \times 10^{43}$& 0.3--1 &  Median luminosity \\[4pt] 

ASASSN-14ae\cite{Chornock14} & 10--100& $<1 \times 10^{41}$ &0.3--10  & Co-added XRT data \\[4pt] 

PS1-10jh\cite{Gezari12} & 365 & $<6\times 10^{41} $&0.1--10 &  \\ [4pt] 

D1-9\cite{Gezari09}  			& 840 & \phantom{$<$} $4\times 10^{41}$& $\approx 0.2$ & Four photons detected \\[4pt] 

J0952􏰈+21\cite{Komossa09} & $\sim 10^3$ & $1 \times 10^{41}$ & 0.1--10 & Start of flare not constrained, \\
J0952􏰈+21  & $\sim 10^3$ & $4 \times 10^{40}$ & 10--20 &  relatively hard x-ray spectrum \\[4 pt]

D23-H1\cite{Gezari08}   	& 3 & $<7\times 10^{40}		$& 0.2--2& \\
D23-H1  	& 116 & $<1\times 10^{41}$& 0.2--10&   \\[4pt]

D3-13\cite{Gezari08}		& 437 & $<2\times 10^{42}$& 0.3--2  &  \\ 
D3-13 	& 437+1 & \phantom{$<$} $2\times 10^{43}$& 0.3--1 &   After previous non-detection \\
D3-13	& 437+30 & $<1\times 10^{42}$&0.3--2 & After first observation \\
\hline
\end{tabular}
\end{center}
\end{table}

Up to now, $\approx 12$ nuclear flares that are strong TDF candidates have been found in optical/UV imaging surveys (including ASASSN-14li), which all share two key properties: blue colors (Fig.~\ref{fig:uggr}) and a constant black body temperature (table~\ref{tab:lc}). The mean K-corrected $g$-band peak absolute magnitude of these 12 sources is $m_g= -18.9$ and the scatter is 1.0~mag. With an observed peak absolute magnitude of $m_g=-17.7$, \mbox{ASASSN-14li} is at the faint end of the TDF population but not an outlier. The light curve of \mbox{ASASSN-14li} appears similar to other well-sampled TDFs such as PS1-10jh \cite{Gezari12} and PTF-09ge \cite{Arcavi14} (Fig.~\ref{fig:loglin}). We note that TDFs can also be identified in spectroscopic surveys \cite{Komossa08} and a few strong candidates have been reported \cite{Komossa09,Wang12}. The peak luminosity of these events is poorly constrained, hence we cannot compare them directly to \mbox{ASASSN-14li}.

About a dozen TDFs have been found in x-ray surveys [a recent review \cite{Komossa15} lists 13 high amplitude x-ray flares, which is a subsample of all x-ray TDFs, selected to have an established galaxy counterpart without AGN signatures in its optical spectrum]. The x-ray selected TDFs have ``soft'' spectra at the time of their peak. The best-studied thermal x-ray TDF, NGC~5905 \cite{Bade96,KomossaBade99}, has a luminosity of $L_X= 7 \times 10^{42}\,{\rm erg}\,{\rm s}^{-1}$ and $kT=0.06$~keV at its peak, both of which are similar to the x-ray properties of \mbox{ASASSN-14li}. The x-ray light curve of ASASSN-14li is difficult to compare to past soft x-ray-selected TDF candidates, because it is by far the best sampled.  Past TDF candidates found through thermal x-ray emission show decaying light curves that can often be fit to a $t^{-5/3}$ power law, but the cadence of these observations is typically $\sim 6$ months or even longer \cite{KomossaBade99,Esquej08}. At late times, the light curve of \mbox{ASASSN-14li}, appears consistent with a $t^{-5/3}$ decay. 

Six tidal flares found by their UV or optical signatures have been followed-up with x-ray observations (table~\ref{tab:xfol}). Five of these were either not detected, or were two orders of magnitude fainter than \mbox{ASASSN-14li}. The sixth tidal disruption flare, D3-13, was detected at $L_X\sim 10^{43}\,{\rm erg}\,{\rm s}^{-1}$, but only in one out of three {\it Chandra} exposures, implying a factor of 10 variability on day timescales, a large difference with the x-ray light curve of \mbox{ASASSN-14li}.

Besides the two dozen TDFs that have been found through their thermal emission, three more TDFs have been found their non-thermal (hard x-ray) emission triggering the {\it Swift} Burst Alert Telescope (BAT) \cite{Bloom11, Levan11,Burrows11, Cenko11, Brown15}. The non-thermal emission is explained by a newly-launched jet with a small inclination to our line of sight \cite{Bloom11}.
Thermal emission from these events has been found in sensitive optical/infra-red follow-up observations \cite{Cenko11,Pasham15,Brown15} --- although a caveat is optical synchrotron emission, which can also explain part of these observations, as evidenced by the relatively high degree of optical polarization seen for {\it Swift}~J1644+57 \cite{Wiersema12}. The peak black body luminosity of {\it Swift}~J2058+05 is $\sim 10^{44.7}\,{\rm erg}\,{\rm s}^{-1} $ and the total energy emitted over 176~days is $\sim10^{51.7}$~erg \cite{Pasham15}. This can be compared to the peak black body luminosity of \mbox{ASASSN-14li} ($\sim 10^{43.7}\,{\rm erg}\,{\rm s}^{-1}$ and $\sim 10^{43.8}\,{\rm erg}\,{\rm s}^{-1}$ for the UVOT and XRT data, respectively) and total black body energy emitted ($10^{50.5}+10^{50.9}=10^{51.0}$~erg, for the sum of the UVOT and XRT data, respectively). The ratio of the integrated thermal luminosity and the total jet energy is $\sim 10^{0}$ and $\sim 10^{2.5}$ for {\it Swift}~J2058+05 \cite{Pasham15} and \mbox{ASASSN-14li}, respectively. This large difference in jet efficiency supports our hypothesis that the non-thermal TDFs selected by {\it Swift} BAT observations are a different class of jets.

Radio observations of previous thermal TDF candidates found prior to 2013 are summarized in \cite{vanVelzen12b}. In table~\ref{tab:predict} we present an updated compilation using more recent radio upper limits \cite{Chornock14,Arcavi14}. 
In general, the radio flares predicted by our model for jets with $E\sim10^{48}$~erg are an order of magnitude below the flux limit of follow-up observations for almost all previously observed TDF candidates; the one exception is NGC~5905, which would be marginally detectable (although the predicted flux is uncertain since the radio observations were obtained 6~years after the peak of this TDF where our model predictions are likely less accurate).

We note that the Bower et~al. (2013) \cite{Bower13} sample of Karl~G.~Jansky VLA (JVLA) follow-up observations of x-ray selected thermal TDF candidates contains two detections with a flux level of $\sim 100$~$\mu$Jy, both obtained about 20 years after the x-ray detection. However, both radio detections are unlikely to originate from TDFs. The first, IC~3599 \cite{Grupe95}, is an optically variable Seyfert galaxy that recently showed a second x-ray outburst, all of which imply a long-lived AGN as the source of the x-ray flare \cite{Grupe15} and the observed radio emission \cite{Komossa09}. Bower et~al. (2013) also detected radio emission within the error circle of the x-ray flare RX~J1420+5334 \cite{Greiner00}, but the emission does not coincide with the low-redshift galaxy (2MASS J14202436+5334117; $z=0.147$) that is the most likely source of the x-ray flare. Instead the source of the radio emission is a higher redshift source (SDSS J142025.18+533354.9; $z=0.522$), found at the edge of the 20" circle that denotes the 3-$\sigma$ uncertainty of the ROSAT all-sky survey detection. Since the median redshift of the other ROSAT-selected TDFs is 0.037,  the radio-emitting galaxy at $z=0.522$ is almost certainly a background source, i.e., unrelated to the x-ray flare. Finally, we note that both radio detections (IC~3599 and SDSS J142025.18+533354.9) are not confirmed transients or variables, since they have been observed at JVLA-sensitivity only once. 

\begin{table}[t]
\small
\begin{center}
\caption{{\bf Predicted flux of ASASSN-14li at the redshift of other thermal TDFs}. The fourth column ($\Delta T$) denotes the rest-frame time since maximum light. We show the $5 \sigma$ upper limit of the radio flux of known TDFs and the predicted flux of ASASSN-14li in the rest-frame and observing frequency of the corresponding thermal TDF as extrapolated using our jet model.}
\label{tab:predict}
\begin{tabular}{lccccc}
\hline 
Name & Redshift & $\nu$ & $\Delta T$ & $F_{\nu}$   & $F_{\nu, \rm predicted}$  \\
&  &  (GHz) & (yr)          & ($\mu$Jy)  & ($\mu$Jy) \\
\hline \hline
ASASSN-14li 					&0.021   &  15.7		& 0.08  &   $ (1.9 \pm 0.1)\times 10^{3}$   & $2.1 \times 10^{3}$ \\ 
\hline
PS1-11af\cite{Chornock14}    			&0.405  	&    4.9  &  0.13   &    $<85$                		&     3.8       \\
PS1-10jh\cite{vanVelzen12b}      			&0.170   	&    5.0  	&  0.49   &    $<75$                &   	 8.2       \\
SDSS-TDE2\cite{Bower13}        &0.252  	&    8.4  &  1.1    &    $<100$               		   &     1.7     \\
SDSS J1201+30\cite{Saxton12} &0.146  	&    4.8  &  1.4    &    $<220$               		   &     5.7       \\
PTF-10iya\cite{vanVelzen12b}      		&0.224  	&    5.0  &  1.6    &    $<40$                		   &     1.9        \\
GALEX-D3-13\cite{vanVelzen12b}     &0.370   	&    1.4  &  1.8    &   $<150$               		   &     0.6       \\
PS1-11af\cite{Chornock14}      			&0.405  	&    5.5  &  1.0   &    $<50$                		   &     0.8       \\
PS1-11af\cite{Chornock14}      			&0.405  	&    5.5  &  2.4   &    $<75$                		   &     0.3       \\
SDSS-TDE1\cite{vanVelzen12b}        &0.136  	&    5.0  &  5.4    &    $<50$                	   &     1.6       \\
SDSS-TDE2\cite{vanVelzen12b}        &0.252  	&    5.0  &  4.3    &    $<60$                		   &     0.5       \\
GALEX-D23-H1\cite{vanVelzen12b}   &0.186  	&    5.0  &  4.8    &    $<40$                		   &     0.9       \\
PTF-09axc\cite{Arcavi14}  		&0.115 	&    3.5  &  5.0   &    $<550$               		   &     2.9      \\
PTF-09axc\cite{Arcavi14}      		&0.115 	&    6.1  &  5.0   &    $<250$               		   &     2.2      \\
NGC 5905\cite{Komossa02}      		&0.011  	&    8.5  &  6.0    &    $<150$               		   &     184     \\
GALEX-D3-13\cite{vanVelzen12b}     &0.370  	&    5.0  &  7.6    &    $<40$                		   &     1.1       \\
GALEX-D1-9\cite{vanVelzen12b}       &0.326  	&    5.0  &  8.0    &    $<45$                		   &     0.1       \\
SDSS J1311--01\cite{Bower13} &0.195  	&    3.0  &  8.4    &    $<95$                		   &     0.4       \\
SDSS J1323+48\cite{Bower13} &0.086 	&    3.0  &  8.6    &    $<170$               		   &     3.2       \\
RX J1242--1119\cite{Bower13} &0.050   	&    3.0  &  20   &    $<90$                		   &     4.2       \\
RX J1624+7554\cite{Bower13} 	&0.064  	&    3.0  &  22   &    $<85$               		   &     2.0       \\
\hline
\end{tabular}
\end{center}
\end{table}

%
%
\section{Origin of radio emission}\label{sec:radio_theory}
%
%

In this section we first establish that the post-flare radio emission from ASASSN-14li is unlikely to be dominated by the jet that existed prior to the flare and its origin is unlikely to be a non-relativistic gas flow. We then proceed to model the observed emission as synchrotron radiation from a spherical blast wave that was created after the disruption. 

\subsection{Pre-flare radio emission}
The radio emission from radiatively inefficient AGN manifests itself as a flat spectrum radio core \cite{Nagar00,Falcke01}, implying a compact conical radio jet \cite{Blandford79}. For a flat baseline radio SED ($\alpha=0$), the AMI 15.7~GHz light curve presents a factor 10 decrease with respect to this baseline. Further evidence for the termination of the original jet is the decrease of the 1.4~GHz flux with respect to the baseline flux from FIRST.  Variability at 1.4~GHz is rare;  only 0.1\% of compact radio sources show variability at the level observed for \mbox{ASASSN-14li} \cite{Ofek11b}. Most of this variability is by induced scintillation, which can occur for both high- and low-redshift jets \cite{Gaensler00}.

With the termination of the central engine, there is no longer a source of energy for particle acceleration upstream in the jet. The synchrotron luminosity of the pre-flare AGN jet is thus determined by an aging population of electrons that is traveling radially away from the black hole with a bulk Lorentz factor $\Gamma_{\rm AGN}\approx 4$ \cite{Falcke01}. Since the total magnetic energy is conserved, the conical jet geometry implies $B\propto R^{-1}$. As the last relativistic electrons stream down the conical jet, the luminosity in the jet-frame decreases as  $L_\nu \propto B^{-7/2} R^2 \propto R^{-3/2}\propto t^{-3/2}$. 

The timescale for the flux decrease due to the diluting magnetic energy density is set by the light travel time to the location where the original jet becomes optically thin to synchrotron self-absorption \cite{Blandford79,Falcke95I,vanVelzen11}. Very Long Baseline Interferometry (VLBI) observations of NGC~4258 at 22~GHz \cite{Herrnstein97} place this radius at 4000~Schwarzschild radii from the central black hole. The AGN in NGC~4258 and ASASSN-14li have both radiatively inefficient flows with a similar narrow line [OIII] luminosity. We thus find
\begin{equation}
R_{\rm ssa}(\nu) \sim 4\times 10^{15}\,{\rm cm}~ \left(\frac{\nu}{\rm 22\,GHz}\right)^{-1} \frac{M_{\rm BH}}{10^{6.5}M_\odot} \quad.
\end{equation}
Hence the location where most of the 1.4~GHz emission of the original jet was produced is $R_{\rm ssa}\sim 10^{17}\,{\rm cm}$ or a few light months. A factor two decrease of the 1.4~GHz flux in the observer frame could occur on the timescale of $\sim  2^{2/3} R_{\rm ssa}\Gamma_{\rm j}/c \sim 10^2$~days, while at 16~GHz the decrease will be twice as fast. 

A second process that will decrease the radio emission of the original jet is inverse Compton (IC) cooling of the electrons by the photons produced in the tidal flare. As evident from Fig.~1 of the main text, the tidal disruption drastically increases the photon energy density near the black hole. The IC cooling time can be estimated to be \cite{Kumar13}:
\begin{equation}\label{eq:tIC}
\tau_{\rm IC} = \frac{3}{4}\frac{ \, m_e c^2}{\sigma_T U_{\rm ph}' \gamma_e} \sim 10~{\rm days} \left(\frac{\theta_{\rm j}}{0.1}\right)^2 \left(\frac{L_{\rm TDF}}{10^{44.6}{\rm erg}\,{\rm s}^{-1}}\right)^{-1} \left(\frac{\Gamma_{\rm j}}{4}\right)^2 \left(\frac{\nu}{1.4\,{\rm GHz}}\right)^{-15/7}
\end{equation}
with $U_{\rm ph}'$ the energy density of the radiation field in the frame of the jet, $\theta_{\rm j}$ the opening angle of the jet (which determines the radial scale of the region that is being cooled, $\theta_{\rm j} R_{\rm ssa}$), $\sigma_T$ the Thomson scattering cross section, $m_e$ the electron mass, and $\gamma_e$ the Lorentz factor of the electrons that emit the bulk of the observed synchrotron luminosity. The latter is estimated using the equipartition magnetic field, which is well-justified by observations of flat spectrum compact jet cores \cite{Sokolovsky11}. The left hand side of Eq.~\ref{eq:tIC} is a general expression, while the right hand side has been normalized to the observed properties of \mbox{ASASSN-14li}. The total luminosity of the seed photons is normalized to the Eddington limit, which is a factor ten larger than the peak of the observed optical/UV luminosity. Due to the unknown contribution at extreme-UV frequencies and dust extinction in the host galaxy, a factor ten bolometric correction is not uncommon for TDFs [cf. \cite{Gezari09}, their Fig.~7]. We conclude that IC cooling will rapidly remove any flux from the pre-existing jet at $\nu>10$~GHz, while at 1.4~GHz the flux decrease could be a factor few at the time of the first WSRT observation. 

Inverse Compton cooling has no effect on  synchrotron emission at large ($R>10^{17}$~cm) scales. Constraining the contribution from a diffuse, optically thin ($\alpha\approx=-0.8$) component to the post-flare light curve would be possible with late-time data (after transient emission has faded), VLBI observations (to measure the fraction of compact flux) or low-frequency radio observations.

\subsection{Non-relativistic outflows}
The transrelativistic blast wave required to power the observed radio emission likely arises from a decelerating relativistic jet.  It cannot be produced by the dynamics of the disruption event itself, because typical velocities of the unbound debris are $v_{\rm u}\sim \sqrt{\Delta \epsilon}$, where the spread in debris specific energy is $\Delta \epsilon = GM_{\rm BH} R_\star/R_{\rm T}^2$ \cite{Stone+13}, with $R_{T}$ the tidal disruption radius. For a solar-type star and $M_{\rm BH} = 10^6M_\odot$, this gives a very subrelativistic $v \sim 10^8~{\rm cm~s}^{-1}$.  
If the line widths measured in the optical spectra of TDFs [typically $\sim5000\,{\rm km}\,{\rm s}^{-1}$ \cite{Arcavi14}] are the result of Doppler broadening in an outflow, this outflowing gas is also highly sub-relativistic.

Subrelativistic outflows are unlikely to produce significant radio synchrotron luminosity.  This is because synchrotron self-absorption (SSA) will become important for accelerated particles emitting on too small a radial scale $R=v t$, where $t$ is the elapsed time since the launching of the outflow.  As shown in Section~1, the observed spectral index between 1.4~GHz and 16~GHz is $\alpha =-0.4$ and $\alpha=-0.6$. If we use the first WSRT epoch to put a conservative upper limit on the non-transient flux at 1.4~GHz, we find $\alpha>0$. We can thus conclude that at $\nu \approx 10$~GHz the synchrotron spectrum is optically thin.  

To avoid SSA, the brightness temperature must remain below $\sim6\times 10^9~{\rm K}$.  We can estimate the brightness temperature of the emitting source as
\begin{equation}
T_{\rm B}(\nu) = \frac{c^4}{2k_{\rm B}\nu^2} \frac{S_{\nu}d^2}{4\pi R^2} =  \frac{S_{\nu} d^2}{8 \pi k_{\rm B}\nu^2 v^2 t^2}.
\end{equation}
The first equality is general; the second assumes the emitted radiation originates from the front of an outflow.  At $t=100~{\rm days}$ and $\nu=10$~GHz,  emission from particles in a $v<10^9\,{\rm cm}\,{\rm s}^{-1}$ outflow implies a brightness temperature of  $T_{\rm B} > 7 \times 10^{10}~{\rm K}$. For non-relativistic velocities, the observed radio emission would therefore be well into the synchrotron self-absorption regime ($\alpha>2$), which is clearly inconsistent with the spectral index between 1.4~GHz and 16~GHz. Particle acceleration at small radii in, e.g., circularization shocks, can likewise be ruled out.

\subsection{Blast wave model}
To estimate the radio emission from a spherical blast wave created after the disruption we follow the formalism of \cite{Nakar&Piran11}. The blast wave could be the result of an initially collimated jet that has swept up enough gas in the circumnuclear medium to spread laterally and become approximately spherical \cite{Granot+02,VanEerten+10,Wygoda+11,Giannios11}.  Though clearly a simplification, a spherical model with appropriately chosen energy and external density does a surprisingly good job of reproducing the late-time  radio data from {\it Swift} J1644+57 compared to the results of detailed multi-dimensional hydrodynamical simulations of the deceleration of tidal disruption jets \cite{Mimica15}.  Since at late times the radio emitting plasma is expanding at mildly relativistic velocities and hence is roughly isotropic, a spherical blast wave should reasonably reproduce the radio emission as observed by an off-axis observer and is a reasonable starting point for an analysis given our limited knowledge of the geometry of the system relative to our line of sight.    

Figure~2 of the main text shows a comparison between a model fit and the radio light curves. We assume that fractions of $\epsilon_{e} = 0.2$ and $\epsilon_{B} = 0.01$ of the post-shock energy are used to accelerate relativistic electrons and to amplify the magnetic field, respectively.  We make the standard assumption that relativistic electrons are accelerated into a power-law distribution $N(E) \propto E^{-p}$, with a power-law index $p = 2.1$, similar to that used to model gamma-ray burst afterglows \cite{Panaitescu&Kumar01}. The blast wave is assumed to begin with a mildly relativistic velocity (initial Lorentz factor $\Gamma = 2$), to possess a total energy of $E \approx 2\times 10^{48}$ erg, and to propagate into a medium of density $n = 1000$ cm$^{-3}$ about 20 days prior to the first {\it Swift} observation. Our results are not strongly dependent on the assumed start time of the circumnuclear medium-jet interaction. Since we assume a point injection of energy, the earliest phases of the interaction, before the blast wave has reached a self similar evolution, cannot be precisely defined by our model.

The 15.7~GHz emission peaks at early times, within days of the injection of energy.  However, if the energy were instead injected over an extended interval of time, the early light curve would flatten somewhat.  The delayed rise of the 1.4 GHz flux is due to the fact that the self-absorption frequency remains between the 1.4 and 15.7~GHz bands until it drops below 1.4 GHz at $t \sim 130$ days, after which time the 1.4 GHz flux follows a similar decay as the 15.7~GHz light curve.   

The energy of the blast wave that we infer to be $\sim$ few $\times 10^{48}$ erg is at least 3$-$4 orders of magnitude smaller than the kinetic power of relativistic jets responsible for powering the hard x-ray and radio emission of {\it Swift} J1644+57 \cite{Zauderer11,Berger12,Metzger12,BarniolDuran&Piran13} and {\it Swift}  J2058+05 \cite{Cenko11,Pasham15}.

To decelerate by the time of our first radio observations we require a density of $\sim 10^{3}$ cm$^{-3}$. Is such a high density reasonable in the galactic nucleus on the scales of the jetted emission?  If prior to the TDF the black hole is accreting spherically at a rate of $\dot{M}$, then the density at radius $r$ due to Bondi accretion is given by
\begin{equation}
n \approx \frac{\dot{M}}{4\pi m_p r^{2}v_{\rm ff}} \approx 5000\,{\rm cm^{-3}}\,\left(\frac{\dot{M}}{10^{-3}\dot{M}_{\rm Edd}}\right)\left(\frac{M_{\rm BH}}{10^{6}M_{\odot}}\right)^{1/2}\left(\frac{r}{10^{16}\,{\rm cm}}\right)^{-3/2}
\label{eq:n}
\end{equation}
where $v_{\rm ff} = (2GM_{\rm BH}/r)^{1/2}$ is the free-fall velocity and we have normalized $\dot{M}$ to the Eddington accretion rate $\dot{M}_{\rm Edd} \equiv L_{\rm Edd}/\eta c^{2}$ for a fiducial radiative efficiency $\eta = 0.1$.  

The jet of half-opening angle $\theta \approx 0.1$ and initial Lorentz factor $\Gamma \approx 2$ will decelerate significantly and spread sideways (assuming it has not already done so at an earlier stage; see below) once it has swept up a gaseous mass $M_{\rm swept} \sim (\pi \theta^{2}/3)m_p n r^{3}$ which is comparable to its own rest mass $M_{\rm j} \approx E/\Gamma c^{2}$.  This approximately occurs at the radius
\begin{equation}
r_{\rm dec} \approx 4\times 10^{16}\,{\rm cm}~\Gamma^{-2/3}\left(\frac{E}{10^{48}\,{\rm erg}}\right)^{2/3}\left(\frac{\theta}{0.1}\right)^{-4/3}\left(\frac{\dot{M}}{10^{-3}\dot{M}_{\rm Edd}}\right)^{-2/3}\left(\frac{M_{\rm BH}}{10^{6}M_{\odot}}\right)^{-1/3}.
\end{equation}
For values of $\Gamma \sim $ few and $E \sim $ few $10^{48}$ erg (as needed to fit the radio data), and for the ranges of accretion rates $\dot{M} \lesssim 10^{-3}\dot{M}_{\rm Edd}$ that characterize the pre-flare low luminosity AGN  (see Section~\ref{sec:host}), we find $r_{\rm dec} \sim 10^{17}$ cm assuming a narrowly collimated jet of $\theta = 0.1$ similar to those of AGN jets. The density at such a radius from equation (\ref{eq:n}) is $n(r_{\rm dec}) \sim 600$ cm$^{-3}$, in reasonable agreement with the value of $\sim 10^{3}$ cm$^{-3}$ needed to fit the radio light curves.

Deceleration of the jet would occur more rapidly if it precesses due to Lense-Thirring torques from misaligned BH spin \cite{Stone&Loeb12}.  Given a misalignment angle $\psi$ between the jet axis and the BH spin axis, $M_{\rm swept}$ increases by a factor up to $2\sin\psi / \theta$ and $r_{\rm dec}$ decreases by a factor up to $(4\sin\psi / \theta)^{-2/3}$ (the enhancement will be smaller if the jet precession timescale is much longer than $r_{\rm dec}/c$, or the misalignment angle is very small).

A significantly wider initial angle could also arise naturally in a weak jet. As shown in \cite{Bromberg+11}, a hydrodynamical jet only remains collimated by its environment of density $n$ if its luminosity exceeds a critical value of 
\begin{eqnarray}
L_{\rm j,min} &\approx&  n m_p c^{3} r^{2} \theta^{5/3}  \nonumber \\
&\approx& 5\times 10^{41}\,{\rm erg\,s^{-1}} \left(\frac{\theta}{0.1}\right)^{5/3}\left(\frac{\dot{M}}{10^{-3}\dot{M}_{\rm Edd}}\right)\left(\frac{M_{\rm BH}}{10^{6}M_{\odot}}\right)^{1/2}\left(\frac{r}{10^{16}\,{\rm cm}}\right)^{1/2}, \nonumber \\[-10pt]
\label{eq:Lcollimate}
\end{eqnarray}
where we have used equation (\ref{eq:n}) to substitute $n$.  If the total inferred jet energy of $E \sim$ few $10^{48}$ erg were released over a duration of $t_{\rm j} \sim $ month, then the luminosity of $L_{\rm j} \sim E/t_{\rm j} \sim 10^{42}$ erg s$^{-1}$ is comparable to the limit of equation (\ref{eq:Lcollimate}).  The jet could thus have begun highly collimated near the black hole, but was disrupted at larger radii ($L_{\rm j,min} \propto r^{1/2}$).

Finally, we expand here on an implication of the host galaxy's prior AGN.  The circumnuclear medium densities required to explain our radio light curves are easily attainable in a moderately sub-Eddington Bondi flow, but are orders of magnitude higher than what would exist in the cone of a pre-existing jet.  In order to explain observed emission from the tidal disruption jet, the jet must either (i) have a significantly larger $\theta$ than did the AGN jet, or (ii) they must be pointed along different axes.  The likelihood of the first possibility is difficult to quantify without observations of the rise of the radio light curve; the second would imply that jets from misaligned accretion disks are generally launched along the disk angular momentum vector, not the BH spin vector, in tension with the analysis of {\it Swift} J1644+57 presented in \cite{Stone&Loeb12}.

\begin{table}
\caption{{\bf Black body parameters from UVOT and XRT observations}}\label{tab:bb}
\begin{center}
\tiny
\begin{tabular}{c|ccc|cccc}
& \multicolumn{3}{c|}{UVOT} & \multicolumn{3}{c}{XRT}\\
Epoch  & $T$ & $R$ & $\chi^2/{\rm dof}$ & $T$  &$R$ & $\chi^2/{\rm dof}$ \\ 
(MJD)  & ($\times 10^{4} $~K) & ($\times 10^{14}$~cm) & & ($\times 10^{5}$~K) & ($\times 10^{11}$~cm) & \\
 \hline

$56991.4$ & $3.5\pm0.3$ & $2.2\pm0.2$ & $4.2/4$ & $7.7\pm0.2$ & $5.0\pm0.4$ & $64.3/60$\\
$56993.9$ & $2.6\pm0.2$ & $2.7\pm0.3$ & $17.4/4$ & $7.4\pm0.2$ & $5.9\pm0.5$ & $72.3/63$\\
$56995.3$ & $3.8\pm0.4$ & $2.0\pm0.2$ & $6.5/4$ & $7.7\pm0.2$ & $4.5\pm0.4$ & $72.2/58$\\
$56998.2$ & $3.6\pm0.4$ & $2.0\pm0.2$ & $9.8/4$ & $8.0\pm0.2$ & $4.9\pm0.4$ & $91.0/65$\\
$57001.6$ & $3.5\pm0.4$ & $2.1\pm0.2$ & $5.3/4$ & $7.9\pm0.2$ & $4.7\pm0.4$ & $92.1/63$\\
$57004.1$ & $3.8\pm0.5$ & $1.8\pm0.2$ & $3.8/4$ & $7.5\pm0.3$ & $5.7\pm0.8$ & $59.9/50$\\
$57007.3$ & $3.7\pm0.5$ & $1.7\pm0.2$ & $4.9/4$ & $8.0\pm0.1$ & $4.9\pm0.3$ & $104.4/75$\\
$57010.8$ & $3.5\pm0.4$ & $1.8\pm0.2$ & $6.5/4$ & $7.9\pm0.1$ & $5.1\pm0.3$ & $68.6/67$\\
$57013.1$ & $3.4\pm0.4$ & $1.9\pm0.2$ & $9.9/4$ & $8.0\pm0.1$ & $4.6\pm0.3$ & $83.2/74$\\
$57016.1$ & $3.5\pm0.4$ & $1.7\pm0.2$ & $7.6/4$ & $7.7\pm0.2$ & $5.3\pm0.5$ & $57.5/65$\\
$57019.5$ & $3.4\pm0.4$ & $1.6\pm0.2$ & $3.5/4$ & $7.7\pm0.1$ & $5.0\pm0.4$ & $119.2/72$\\
$57022.7$ & $3.1\pm0.3$ & $1.8\pm0.2$ & $7.7/4$ & $7.9\pm0.2$ & $4.2\pm0.4$ & $52.0/66$\\
$57029.3$ & $2.9\pm0.4$ & $1.6\pm0.3$ & $7.5/4$ & $8.1\pm0.1$ & $4.3\pm0.3$ & $88.8/77$\\
$57033.0$ & $5.1\pm1.3$ & $1.0\pm0.2$ & $6.1/4$ & $7.6\pm0.2$ & $5.3\pm0.5$ & $63.0/56$\\
$57036.0$ & $2.6\pm0.3$ & $1.9\pm0.3$ & $3.5/4$ & $7.5\pm0.2$ & $5.7\pm0.5$ & $66.4/58$\\
$57039.0$ & $4.1\pm0.7$ & $1.2\pm0.2$ & $1.2/4$ & $7.6\pm0.2$ & $5.1\pm0.4$ & $77.2/58$\\
$57042.2$ & $3.3\pm0.5$ & $1.4\pm0.3$ & $5.1/4$ & $7.9\pm0.2$ & $4.6\pm0.4$ & $51.5/65$\\
$57045.6$ & $3.6\pm0.6$ & $1.2\pm0.2$ & $2.8/4$ & $7.9\pm0.2$ & $4.6\pm0.4$ & $77.3/63$\\
$57048.7$ & $2.9\pm0.4$ & $1.5\pm0.2$ & $2.0/4$ & $7.5\pm0.2$ & $5.2\pm0.5$ & $58.2/55$\\
$57051.5$ & $3.9\pm0.8$ & $1.1\pm0.2$ & $5.0/4$ & $7.2\pm0.2$ & $6.1\pm0.6$ & $53.8/55$\\
$57054.0$ & $3.3\pm0.5$ & $1.3\pm0.2$ & $0.9/4$ & $7.4\pm0.2$ & $5.3\pm0.5$ & $86.7/60$\\
$57057.5$ & $3.0\pm0.4$ & $1.5\pm0.2$ & $4.0/4$ & $7.2\pm0.2$ & $5.6\pm0.5$ & $52.9/58$\\
$57059.8$ & $3.5\pm0.7$ & $1.2\pm0.3$ & $3.8/4$ & $7.5\pm0.2$ & $4.8\pm0.5$ & $83.5/57$\\
$57065.8$ & $3.9\pm0.9$ & $1.1\pm0.3$ & $1.1/4$ & $7.6\pm0.3$ & $4.3\pm0.6$ & $50.8/49$\\
$57068.8$ & $2.3\pm0.2$ & $1.7\pm0.3$ & $1.1/4$ & $7.0\pm0.3$ & $5.7\pm0.8$ & $52.5/46$\\
$57071.7$ & $3.0\pm0.4$ & $1.4\pm0.2$ & $3.0/4$ & $7.1\pm0.2$ & $5.6\pm0.6$ & $71.0/52$\\
$57074.8$ & $3.4\pm0.6$ & $1.1\pm0.2$ & $2.9/4$ & $7.5\pm0.2$ & $4.5\pm0.5$ & $60.0/57$\\
$57077.6$ & $4.2\pm1.0$ & $0.9\pm0.2$ & $3.0/4$ & $6.7\pm0.3$ & $6.2\pm1.0$ & $46.9/41$\\
$57080.9$ & $3.2\pm0.8$ & $1.1\pm0.3$ & $2.1/4$ & $7.0\pm0.2$ & $5.8\pm0.6$ & $102.6/54$\\
$57086.9$ & $3.6\pm0.7$ & $0.9\pm0.2$ & $0.1/4$ & $7.5\pm0.2$ & $4.5\pm0.5$ & $60.4/55$\\
$57089.3$ & $3.9\pm0.9$ & $0.8\pm0.2$ & $2.9/4$ & $7.7\pm0.2$ & $4.4\pm0.5$ & $108.2/58$\\
$57099.4$ & $3.1\pm0.6$ & $1.0\pm0.2$ & $0.8/3$ & $7.0\pm0.3$ & $5.8\pm0.9$ & $80.2/47$\\
$57102.4$ & $2.8\pm0.6$ & $1.1\pm0.3$ & $4.6/3$ & $7.3\pm0.2$ & $4.7\pm0.6$ & $69.5/54$\\
$57105.2$ & -- & -- & -- & $7.4\pm0.6$ & $3.5\pm1.1$ & $50.3/30$\\
$57108.8$ & $4.0\pm1.3$ & $0.7\pm0.2$ & $1.1/3$ & $6.9\pm0.2$ & $5.9\pm0.8$ & $50.6/48$\\
$57111.9$ & $3.1\pm0.6$ & $0.9\pm0.2$ & $3.5/3$ & $7.2\pm0.2$ & $5.3\pm0.6$ & $46.9/53$\\
$57114.1$ & $3.8\pm1.2$ & $0.7\pm0.3$ & $4.2/3$ & $7.4\pm0.3$ & $4.3\pm0.6$ & $74.2/50$\\
$57117.7$ & $3.1\pm0.6$ & $0.9\pm0.2$ & $5.9/3$ & $7.6\pm0.3$ & $3.8\pm0.5$ & $81.0/56$\\
$57120.3$ & $2.5\pm0.4$ & $1.1\pm0.3$ & $1.5/3$ & $7.2\pm0.3$ & $4.5\pm0.7$ & $73.1/42$\\
$57123.5$ & $3.0\pm0.6$ & $0.8\pm0.2$ & $1.8/3$ & $7.2\pm0.2$ & $4.7\pm0.6$ & $46.8/45$\\
$57126.1$ & $3.2\pm0.9$ & $0.7\pm0.3$ & $1.6/3$ & $7.1\pm0.3$ & $4.7\pm0.6$ & $43.4/49$\\
$57129.2$ & $3.3\pm1.1$ & $0.7\pm0.3$ & $0.0/3$ & $7.0\pm0.3$ & $5.2\pm0.8$ & $60.6/43$\\
$57132.5$ & $3.8\pm1.5$ & $0.6\pm0.3$ & $5.5/4$ & $7.5\pm0.3$ & $3.8\pm0.5$ & $28.7/45$\\
$57136.1$ & $4.6\pm2.4$ & $0.5\pm0.3$ & $3.0/3$ & $6.7\pm0.4$ & $5.2\pm1.1$ & $48.0/35$\\
$57139.3$ & $5.1\pm2.2$ & $0.5\pm0.2$ & $1.6/3$ & $6.7\pm0.3$ & $6.1\pm1.0$ & $63.2/39$\\
$57147.6$ & $4.9\pm2.2$ & $0.5\pm0.2$ & $3.1/3$ & $6.9\pm0.3$ & $4.5\pm0.7$ & $55.9/46$\\
$57150.2$ & $3.4\pm0.8$ & $0.7\pm0.2$ & $1.8/3$ & $6.5\pm0.2$ & $6.2\pm0.9$ & $47.5/43$\\
$57153.4$ & $2.9\pm0.6$ & $0.8\pm0.2$ & $1.8/3$ & $6.4\pm0.2$ & $6.3\pm1.0$ & $55.1/49$\\
$57156.3$ & $5.6\pm4.6$ & $0.4\pm0.3$ & $2.7/3$ & $7.0\pm0.3$ & $4.6\pm0.8$ & $61.9/40$\\
$57173.0$ & $3.2\pm1.0$ & $0.7\pm0.3$ & $4.1/3$ & $6.7\pm0.3$ & $5.2\pm0.9$ & $36.8/41$\\
$57176.1$ & $2.6\pm0.5$ & $0.9\pm0.3$ & $2.0/3$ & $6.1\pm0.3$ & $7.2\pm1.4$ & $37.6/38$\\
$57179.0$ & $3.3\pm1.6$ & $0.7\pm0.4$ & $1.7/3$ & $6.8\pm0.6$ & $4.5\pm1.4$ & $40.6/25$\\
$57182.1$ & $2.8\pm1.2$ & $0.8\pm0.5$ & $1.0/3$ & $6.6\pm0.5$ & $4.8\pm1.4$ & $32.4/33$\\
$57186.0$ & $4.8\pm2.9$ & $0.4\pm0.2$ & $6.2/3$ & $6.4\pm0.4$ & $5.9\pm1.4$ & $33.1/35$\\
$57188.5$ & -- & -- & -- &  -- & -- & -- &  \\
$57191.8$ & $3.4\pm1.1$ & $0.6\pm0.2$ & $0.4/3$ & $5.8\pm0.3$ & $9.3\pm1.9$ & $59.8/37$\\
$57194.7$ & $3.9\pm2.0$ & $0.5\pm0.3$ & $3.1/3$ & $6.0\pm0.3$ & $7.2\pm1.3$ & $51.3/41$\\
$57200.4$ & $2.9\pm0.7$ & $0.7\pm0.2$ & $8.9/3$ & $5.9\pm0.4$ & $8.0\pm2.0$ & $19.2/30$\\
$57203.8$ & $5.5\pm4.8$ & $0.4\pm0.3$ & $8.7/3$ & $7.1\pm0.4$ & $3.6\pm0.7$ & $29.9/39$\\
$57226.2$ & $6.0\pm5.1$ & $0.4\pm0.3$ & $2.1/3$ & $5.8\pm0.4$ & $7.9\pm2.3$ & $59.1/30$\\
$57230.4$ & $6.6\pm6.1$ & $0.3\pm0.2$ & $1.8/3$ & $6.0\pm0.4$ & $7.3\pm2.2$ & $16.5/25$\\
$57236.4$ & -- & -- & -- &  $6.2\pm0.3$ & $6.1\pm1.3$ & $34.5/34$\\
$57238.8$ & -- & -- & -- &  $6.1\pm0.3$ & $6.5\pm1.4$ & $47.8/33$\\
$57242.1$ & -- & -- & -- &  $6.2\pm0.3$ & $6.1\pm1.3$ & $40.8/33$\\
$57246.9$ & -- & -- & -- &  $6.4\pm0.4$ & $4.8\pm1.1$ & $65.0/37$\\
\end{tabular}
\end{center}
\end{table}


\clearpage



\begin{thebibliography}{100}
\small
\bibitem{Giannios11}
D.~{Giannios}, B.~D. {Metzger}, {Radio transients from stellar tidal disruption
  by massive black holes}.
\newblock {\it Mon. Not. R. Astron. Soc.\/} {\bf 416}, 2102 (2011).

\bibitem{vanVelzen11}
S.~van Velzen, E.~K\"ording, H.~Falcke, Radio jets from stellar tidal
  disruptions.
\newblock {\it Mon. Not. R. Astron. Soc.\/} {\bf 417}, L51 (2011).

\bibitem{Bade96}
N.~{Bade}, S.~{Komossa}, M.~{Dahlem}, {Detection of an extremely soft x-ray
  outburst in the HII-like nucleus of NGC 5905.}
\newblock {\it Astron. Astrophys.\/} {\bf 309}, L35 (1996).

\bibitem{Gezari09}
S.~{Gezari}, {\it et~al.\/}, {Luminous thermal flares from quiescent
  supermassive black holes}.
\newblock {\it Astrophys. J.\/} {\bf 698}, 1367 (2009).

\bibitem{vanVelzen10}
S.~{van Velzen}, {\it et~al.\/}, {Optical discovery of probable stellar tidal
  disruption flares}.
\newblock {\it Astrophys. J.\/} {\bf 741}, 73 (2011).

\bibitem{Bloom11}
J.~S. {Bloom}, {\it et~al.\/}, {A possible relativistic jetted outburst from a
  massive black hole fed by a tidally disrupted star}.
\newblock {\it Science\/} {\bf 333}, 203 (2011).

\bibitem{Levan11}
A.~J. {Levan}, {\it et~al.\/}, {An extremely luminous panchromatic outburst
  from the nucleus of a distant galaxy}.
\newblock {\it Science\/} {\bf 333}, 199 (2011).

\bibitem{Burrows11}
D.~N. {Burrows}, {\it et~al.\/}, {Relativistic jet activity from the tidal
  disruption of a star by a massive black hole}.
\newblock {\it Nature\/} {\bf 476}, 421 (2011).

\bibitem{Cenko11}
S.~B. {Cenko}, {\it et~al.\/}, {Swift J2058.4+0516: Discovery of a possible
  second relativistic tidal disruption flare?}
\newblock {\it Astrophys. J.\/} {\bf 753}, 77 (2012).

\bibitem{Brown15}
G.~C. {Brown}, {\it et~al.\/}, {Swift J1112.2-8238: A candidate relativistic
  tidal disruption flare}.
\newblock {\it Mon. Not. R. Astron. Soc.\/} {\bf 452}, 4297 (2015).

\bibitem{Zauderer11}
B.~A. {Zauderer}, {\it et~al.\/}, {Birth of a relativistic outflow in the
  unusual {$\gamma$}-ray transient Swift J164449.3+573451}.
\newblock {\it Nature\/} {\bf 476}, 425 (2011).

\bibitem{Berger12}
E.~{Berger}, {\it et~al.\/}, {Radio monitoring of the tidal disruption event
  Swift J164449.3+573451. I. Jet energetics and the pristine parsec-scale
  environment of a supermassive black hole}.
\newblock {\it Astrophys. J.\/} {\bf 748}, 36 (2012).

\bibitem{Bower13}
G.~C. {Bower}, B.~D. {Metzger}, S.~B. {Cenko}, J.~M. {Silverman}, J.~S.
  {Bloom}, {Late-time radio emission from X-Ray-selected tidal disruption
  events}.
\newblock {\it Astrophys. J.\/} {\bf 763}, 84 (2013).

\bibitem{vanVelzen12b}
S.~{van Velzen}, D.~A. {Frail}, E.~{K{\"o}rding}, H.~{Falcke}, {Constraints on
  off-axis jets from stellar tidal disruption flares}.
\newblock {\it Astron. Astrophys.\/} {\bf 552}, A5 (2013).

\bibitem{Fender04}
R.~P. {Fender}, T.~M. {Belloni}, E.~{Gallo}, {Towards a unified model for black
  hole x-ray binary jets}.
\newblock {\it Mon. Not. R. Astron. Soc.\/} {\bf 355}, 1105 (2004).

\bibitem{Holoien15}
T.~W.-S. {Holoien}, {\it et~al.\/}, {Six months of multi-wavelength follow-up
  of the tidal disruption candidate ASASSN-14li and implied TDE rates from
  ASAS-SN}.
\newblock {\it Mon. Not. R. Astron. Soc.\/}, in press (available at \verb http://arxiv.org/abs/1507.01598 ).

\bibitem{Falcke04}
H.~{Falcke}, E.~{K{\"o}rding}, S.~{Markoff}, {A scheme to unify low-power
  accreting black holes. Jet-dominated accretion flows and the radio/X-ray
  correlation}.
\newblock {\it Astron. Astrophys.\/} {\bf 414}, 895 (2004).

\bibitem{Miller15}
J.~M.~{Miller}~{\it et~al.\/}, {Flows of x-ray gas reveal the disruption of a star by a massive black hole}.
\newblock {\it Nature\/} {\bf 526}, 542 (2015).

\bibitem{Kanner13}
J.~{Kanner}, {\it et~al.\/}, {X-Ray Transients in the Advanced LIGO/Virgo
  Horizon}.
\newblock {\it Astrophys. J.\/} {\bf 774}, 63 (2013).

\bibitem{Gezari13}
S.~{Gezari}, {\it et~al.\/}, {The GALEX Time Domain Survey. I. Selection and
  classification of over a thousand ultraviolet variable sources}.
\newblock {\it Astrophys. J.\/} {\bf 766}, 60 (2013).

\bibitem{Ulmer99}
A.~{Ulmer}, {Flares from the tidal disruption of stars by massive black holes}.
\newblock {\it Astrophys. J.\/} {\bf 514}, 180 (1999).

\bibitem{Guillochon14}
J.~{Guillochon}, H.~{Manukian}, E.~{Ramirez-Ruiz}, {PS1-10jh: The disruption of
  a main-sequence star of near-solar composition}.
\newblock {\it Astrophys. J.\/} {\bf 783}, 23 (2014).

\bibitem{Gezari12}
S.~{Gezari}, {\it et~al.\/}, {An ultraviolet-optical flare from the tidal
  disruption of a helium-rich stellar core}.
\newblock {\it Nature\/} {\bf 485}, 217 (2012).

\bibitem{Shiokawa15}
H.~{Shiokawa}, J.~H. {Krolik}, R.~M. {Cheng}, T.~{Piran}, S.~C. {Noble},
  {General relativistic hydrodynamic simulation of accretion flow from a
  stellar tidal disruption}.
\newblock {\it Astrophys. J.\/} {\bf 804}, 85 (2015).

\bibitem{Maccarone03}
T.~J. {Maccarone}, {Do x-ray binary spectral state transition luminosities
  vary?}
\newblock {\it Astron. Astrophys.\/} {\bf 409}, 697 (2003).

\bibitem{Mimica15}
P.~{Mimica}, D.~{Giannios}, B.~D. {Metzger}, M.~A. {Aloy}, {The radio afterglow
  of Swift J1644+57 reveals a powerful jet with fast core and slow sheath}.
\newblock {\it Mon. Not. R. Astron. Soc.\/} {\bf 450} (2015).

\bibitem{McKinney13}
J.~C. {McKinney}, A.~{Tchekhovskoy}, R.~D. {Blandford}, {Alignment of
  magnetized accretion disks and relativistic jets with spinning black holes}.
\newblock {\it Science\/} {\bf 339}, 49 (2013).

\bibitem{Kelley14}
L.~Z. {Kelley}, A.~{Tchekhovskoy}, R.~{Narayan}, {Tidal disruption and magnetic
  flux capture: Powering a jet from a quiescent black hole}.
\newblock {\it Mon. Not. R. Astron. Soc.\/} {\bf 445}, 3919 (2014).

\bibitem{Fender15}
R.~{Fender}, {\it et~al.\/}, in {\it Advancing Astrophysics with the Square Kilometre Array},  T.~L. {Bourke} {\it et~al.} Eds. (Proceedings of Science, 2014) {chap. 51}. 

\bibitem{vanVelzen14}
S.~{van Velzen}, G.~R. {Farrar}, {Measurement of the Rate of Stellar Tidal
  Disruption Flares}.
\newblock {\it Astrophys. J.\/} {\bf 792}, 53 (2014).

\setcounter{firstbib}{\value{enumiv}}

\end{thebibliography}

\clearpage
\begin{thebibliography}{100}
\small

\setcounter{enumiv}{\value{firstbib}} 

\bibitem{Condon98}
J.~J. {Condon}, {\it et~al.\/}, {The NRAO VLA Sky Survey}.
\newblock {\it Astron. J.\/} {\bf 115}, 1693 (1998).

\bibitem{White97}
R.~L. {White}, R.~H. {Becker}, D.~J. {Helfand}, M.~D. {Gregg}, {A catalog of
  1.4 GHz radio sources from the FIRST survey}.
\newblock {\it Astrophys. J.\/} {\bf 475}, 479 (1997).

\bibitem{helfand15}
D.~J. {Helfand}, R.~L. {White}, R.~H. {Becker}, {The last of FIRST: The final
  catalog and source identifications}.
\newblock {\it Astrophys. J.\/} {\bf 801}, 26 (2015).

\bibitem{Ofek11b}
E.~O. {Ofek}, D.~A. {Frail}, {The structure function of variable 1.4 GHz radio
  sources based on NVSS and FIRST observations}.
\newblock {\it Astrophys. J.\/} {\bf 737}, 45 (2011).

\bibitem{zwart08}
J.~T.~L. {Zwart}, {\it et~al.\/}, {The Arcminute Microkelvin Imager}.
\newblock {\it Mon. Not. R. Astron. Soc.\/} {\bf 391}, 1545 (2008).

\bibitem{Staley15}
T.~D. {Staley}, G.~E. {Anderson}, {AMIsurvey, chimenea and other tools:
  Automated imaging for transient surveys with existing radio-observatories}.
\newblock {\it Astronomy and Computing}, in press (available at \verb http://arxiv.org/abs/1505.08123 ).

\bibitem{staley15a}
T.~D. {Staley}, G.~E. {Anderson}, {AMIsurvey: Calibration and imaging pipeline
  for radio data}.
\newblock {\it in Astrophysics Source Code Library\/} {\bf ascl:1502.017}
  (2015).

\bibitem{perrott13}
Y.~C. {Perrott}, {\it et~al.\/}, {AMI Galactic Plane survey at 16 GHz - I.
  Observing, mapping and source extraction}.
\newblock {\it Mon. Not. R. Astron. Soc.\/} {\bf 429}, 3330 (2013).

\bibitem{jaeger08}
S.~{Jaeger}, {\it ASP Conf. Ser. Astronomical Data Analysis Software and
  Systems XVII\/}, R.~W. {Argyle}, P.~S. {Bunclark}, J.~R. {Lewis}, eds.
  (Astron. Soc. Pac., San Francisco, 2008), vol. 394, p. 623.

\bibitem{staley15b}
T.~D. {Staley}, G.~E. {Anderson}, {chimenea: Multi-epoch radio-synthesis data
  imaging}.
\newblock {\it in Astrophysics Source Code Library\/} {\bf ascl:1504.005}
  (2015).

\bibitem{swinbank15}
J.~D. {Swinbank}, {\it et~al.\/}, {The LOFAR Transients Pipeline}.
\newblock {\it Astronomy and Computing\/} {\bf 11}, 25 (2015).

\bibitem{sault95}
R.~J. {Sault}, P.~J. {Teuben}, M.~C.~H. {Wright}, {\it ASP Conf. Ser.
  Astronomical Data Analysis Software and Systems IV\/}, R.~A. {Shaw}, H.~E.
  {Payne}, J.~J.~E. {Hayes}, eds. (Astron. Soc. Pac., San
  Francisco, 1995), vol.~77, p. 433.

\bibitem{Drake09}
A.~J. {Drake}, {\it et~al.\/}, {First results from the Catalina Real-Time
  Transient Survey}.
\newblock {\it Astrophys. J.\/} {\bf 696}, 870 (2009).

\bibitem{Wright10}
E.~L. {Wright}, {\it et~al.\/}, {The Wide-field Infrared Survey Explorer
  (WISE): Mission description and initial on-orbit performance}.
\newblock {\it Astron. J.\/} {\bf 140}, 1868 (2010).

\bibitem{Lang14}
D.~{Lang}, {unWISE: Unblurred coadds of the WISE imaging}.
\newblock {\it Astron. J.\/} {\bf 147}, 108 (2014).

\bibitem{Skrutskie06}
M.~F. {Skrutskie}, {\it et~al.\/}, {The Two Micron All Sky Survey (2MASS)}.
\newblock {\it Astron. J.\/} {\bf 131}, 1163 (2006).

\bibitem{stoughton02}
C.~{Stoughton}, {\it et~al.\/}, {Sloan Digital Sky Survey: Early data release}.
\newblock {\it Astron. J.\/} {\bf 123}, 485 (2002).

\bibitem{Martin05}
D.~C. {Martin}, {\it et~al.\/}, {The Galaxy Evolution Explorer: A space
  ultraviolet survey mission}.
\newblock {\it Astrophys. J.\/} {\bf 619}, L1 (2005).

\bibitem{Voges99}
W.~{Voges}, {\it et~al.\/}, {The ROSAT all-sky survey bright source catalogue}.
\newblock {\it Astron. Astrophys.\/} {\bf 349}, 389 (1999).

\bibitem{Schlegel98}
D.~J. {Schlegel}, D.~P. {Finkbeiner}, M.~{Davis}, {Maps of dust infrared
  emission for use in estimation of reddening and Cosmic Microwave Background
  radiation foregrounds}.
\newblock {\it Astrophys. J.\/} {\bf 500}, 525 (1998).

\bibitem{Fitzpatrick99}
E.~L. {Fitzpatrick}, {Correcting for the effects of interstellar extinction}.
\newblock {\it Publ. Astron. Soc. Pac.\/} {\bf 111}, 63 (1999).

\bibitem{Yuan13}
H.~B. {Yuan}, X.~W. {Liu}, M.~S. {Xiang}, {Empirical extinction coefficients
  for the GALEX, SDSS, 2MASS and WISE passbands}.
\newblock {\it Mon. Not. R. Astron. Soc.\/} {\bf 430}, 2188 (2013).

\bibitem{Moustakas13}
J.~{Moustakas}, {\it et~al.\/}, {PRIMUS: Constraints on star formation
  quenching and galaxy merging, and the evolution of the stellar mass function
  from $z = 0-1$}.
\newblock {\it Astrophys. J.\/} {\bf 767}, 50 (2013).

\bibitem{Kauffmann03b}
G.~{Kauffmann}, {\it et~al.\/}, {Stellar masses and star formation histories
  for 10$^{5}$ galaxies from the Sloan Digital Sky Survey}.
\newblock {\it Mon. Not. R. Astron. Soc.\/} {\bf 341}, 33 (2003).

\bibitem{Salim07}
S.~{Salim}, {\it et~al.\/}, {UV Star Formation Rates in the Local Universe}.
\newblock {\it Astrophys. J.s\/} {\bf 173}, 267 (2007).

\bibitem{Abazajian09}
K.~N. {Abazajian}, {\it et~al.\/}, {The Seventh Data Release of the Sloan
  Digital Sky Survey}.
\newblock {\it Astrophys. J.s\/} {\bf 182}, 543 (2009).

\bibitem{Bolton12}
A.~S. {Bolton}, {\it et~al.\/}, {Spectral Classification and redshift
  measurement for the SDSS-III Baryon Oscillation Spectroscopic Survey}.
\newblock {\it Astron. J.\/} {\bf 144}, 144 (2012).

\bibitem{Kormendy13}
J.~{Kormendy}, L.~C. {Ho}, {Coevolution (or not) of supermassive black holes
  and host galaxies}.
\newblock {\it Ann. Rev. Astron. Astrophys.\/} {\bf 51}, 511 (2013).

\bibitem{Lackner12}
C.~N. {Lackner}, J.~E. {Gunn}, {Astrophysically motivated bulge-disc
  decompositions of Sloan Digital Sky Survey galaxies}.
\newblock {\it Mon. Not. R. Astron. Soc.\/} {\bf 421}, 2277 (2012).

\bibitem{Condon02}
J.~J. {Condon}, W.~D. {Cotton}, J.~J. {Broderick}, {Radio Sources and star
  formation in the Local Universe}.
\newblock {\it Astron. J.\/} {\bf 124}, 675 (2002).

\bibitem{baldwin81}
J.~A. {Baldwin}, M.~M. {Phillips}, R.~{Terlevich}, {Classification parameters
  for the emission-line spectra of extragalactic objects}.
\newblock {\it Publ. Astron. Soc. Pac.\/} {\bf 93}, 5 (1981).

\bibitem{Heckman05}
T.~M. {Heckman}, A.~{Ptak}, A.~{Hornschemeier}, G.~{Kauffmann}, {The
  relationship of hard x-ray and optical line emission in low-redshift active
  galactic nuclei}.
\newblock {\it Astrophys. J.\/} {\bf 634}, 161 (2005).

\bibitem{Hopkins07}
P.~F. {Hopkins}, G.~T. {Richards}, L.~{Hernquist}, {An observational
  determination of the bolometric quasar luminosity function}.
\newblock {\it Astrophys. J.\/} {\bf 654}, 731 (2007).

\bibitem{heckman04}
T.~M. {Heckman}, {\it et~al.\/}, {Present-day growth of black holes and bulges:
  the Sloan Digital Sky Survey perspective}.
\newblock {\it Astrophys. J.\/} {\bf 613}, 109 (2004).

\bibitem{Jose14}
J.~{Jose}, {\it et~al.\/}, {ASAS-SN Discovery of an Unusual Nuclear Transient
  in PGC 043234}.
\newblock {\it The Astronomer's Telegram\/} {\bf 6777}, 1 (2014).

\bibitem{Poole08}
T.~S. {Poole}, {\it et~al.\/}, {Photometric calibration of the Swift
  ultraviolet/optical telescope}.
\newblock {\it Mon. Not. R. Astron. Soc.\/} {\bf 383}, 627 (2008).

\bibitem{Kingprofile}
\footnotesize {\verb http://www.swift.ac.uk/analysis/xrt/pileup.php   } \small

\bibitem{Arnaud96}
K.~A. {Arnaud}, {\it Astronomical Data Analysis Software and Systems V\/},
  G.~H. {Jacoby}, J.~{Barnes}, eds. (1996), vol. 101 of {\it Astron.
  Soc. Pac. Conference Series\/}, p.~17.

\bibitem{Cash79}
W.~{Cash}, {Parameter estimation in astronomy through application of the
  likelihood ratio}.
\newblock {\it Astrophys. J.\/} {\bf 228}, 939 (1979).

\bibitem{Wstat}
\footnotesize {\verb http://heasarc.gsfc.nasa.gov/docs/xanadu/xspec/manual/XSappendixStatistics.html   } \small


\bibitem{Chornock14}
R.~{Chornock}, {\it et~al.\/}, {The ultraviolet-bright, slowly declining
  transient PS1-11af as a partial tidal disruption event}.
\newblock {\it Astrophys. J.\/} {\bf 780}, 44 (2014).

\bibitem{schneider10}
D.~P. {Schneider}, {\it et~al.\/}, {The Sloan Digital Sky Survey quasar
  catalog. V. Seventh data release}.
\newblock {\it Astron. J.\/} {\bf 139}, 2360 (2010).

\bibitem{Ross12}
N.~P. {Ross}, {\it et~al.\/}, {The SDSS-III Baryon Oscillation Spectroscopic
  Survey: Quasar target selection for data release nine}.
\newblock {\it Astrophys. J.s\/} {\bf 199}, 3 (2012).

\bibitem{Komossa09}
S.~{Komossa}, {\it et~al.\/}, {NTT, Spitzer, and Chandra spectroscopy of
  SDSSJ095209.56+214313.3: The most luminous coronal-line supernova ever
  observed, or a stellar tidal disruption event?}
\newblock {\it Astrophys. J.\/} {\bf 701}, 105 (2009).

\bibitem{Gezari08}
S.~{Gezari}, {\it et~al.\/}, {UV/optical detections of candidate tidal
  disruption events by GALEX and CFHTLS}.
\newblock {\it Astrophys. J.\/} {\bf 676}, 944 (2008).

\bibitem{Komossa08}
S.~{Komossa}, {\it et~al.\/}, {Discovery of superstrong, fading, iron line
  emission and double-peaked balmer lines of the galaxy SDSS
  J095209.56+214313.3: The light echo of a huge flare}.
\newblock {\it Astrophys. J.\/} {\bf 678}, L13 (2008).

\bibitem{Wang12}
T.-G. {Wang}, {\it et~al.\/}, {Extreme coronal line emitters: tidal disruption
  of stars by massive black holes in galactic nuclei?}
\newblock {\it Astrophys. J.\/} {\bf 749}, 115 (2012).

\bibitem{Arcavi14}
I.~{Arcavi}, {\it et~al.\/}, {A continuum of H- to He-rich tidal disruption
  candidates with a preference for E+A galaxies}.
\newblock {\it Astrophys. J.\/} {\bf 793}, 38 (2014).

\bibitem{Komossa15}
S.~{Komossa}, {Tidal disruption of stars by supermassive black holes: Status of
  observations}.
\newblock {\it Journal of High Energy Astrophysics\/} {\bf 7}, 148 (2015).

\bibitem{KomossaBade99}
S.~{Komossa}, N.~{Bade}, {The giant x-ray outbursts in NGC 5905 and IC 3599: 
   Follow-up observations and outburst scenarios}.
\newblock {\it Astron. Astrophys.\/} {\bf 343}, 775 (1999).

\bibitem{Esquej08}
P.~{Esquej}, {\it et~al.\/}, {Evolution of tidal disruption candidates
  discovered by XMM-Newton}.
\newblock {\it Astron. Astrophys.\/} {\bf 489}, 543 (2008).

\bibitem{Pasham15}
D.~R. {Pasham}, {\it et~al.\/}, {A multiwavelength study of the relativistic
  tidal disruption candidate Swift J2058.4+0516 at late times}.
\newblock {\it Astrophys. J.\/} {\bf 805}, 68 (2015).

\bibitem{Wiersema12}
K.~{Wiersema}, {\it et~al.\/}, {Polarimetry of the transient relativistic jet
  of GRB 110328/Swift J164449.3+573451}.
\newblock {\it Mon. Not. R. Astron. Soc.\/} {\bf 421}, 1942 (2012).

\bibitem{Grupe95}
D.~{Grupe}, {\it et~al.\/}, {X-ray outburst of the peculiar Seyfert galaxy IC
  3599}.
\newblock {\it Astron. Astrophys.\/} {\bf 299}, L5 (1995).

\bibitem{Grupe15}
D.~{Grupe}, S.~{Komossa}, R.~{Saxton}, {IC 3599 did it again: A Second outburst
  of the x-ray transient seyfert 1.9 galaxy}.
\newblock {\it Astrophys. J.\/} {\bf 803}, L28 (2015).

\bibitem{Greiner00}
J.~{Greiner}, R.~{Schwarz}, S.~{Zharikov}, M.~{Orio}, {RX J1420.4+5334 -
  another tidal disruption event?}
\newblock {\it Astron. Astrophys.\/} {\bf 362}, L25 (2000).

\bibitem{Saxton12}
R.~D. {Saxton}, {\it et~al.\/}, {A tidal disruption-like x-ray flare from the
  quiescent galaxy SDSS J120136.02+300305.5}.
\newblock {\it Astron. Astrophys.\/} {\bf 541}, A106 (2012).

\bibitem{Komossa02}
S.~{Komossa}, {\it Lighthouses of the Universe\/}, {M.~Gilfanov, R.~Sunyeav, \&
  E.~Churazov}, ed. (2002), pp. 436--442.

\bibitem{Nagar00}
N.~M. {Nagar}, H.~{Falcke}, A.~S. {Wilson}, L.~C. {Ho}, {Radio sources in
  low-luminosity active galactic nuclei. I. VLA detections of compact,
  flat-spectrum cores}.
\newblock {\it Astrophys. J.\/} {\bf 542}, 186 (2000).

\bibitem{Falcke01}
H.~{Falcke}, {\it Reviews in Modern Astronomy\/}, {R.~E.~Schielicke}, ed.
  (2001), vol.~14, pp. 15--51.

\bibitem{Blandford79}
R.~D. {Blandford}, A.~{K{\"o}nigl}, {Relativistic jets as compact radio
  sources}.
\newblock {\it Astrophys. J.\/} {\bf 232}, 34 (1979).

\bibitem{Gaensler00}
B.~M. {Gaensler}, R.~W. {Hunstead}, {Long-term monitoring of Molonglo
  calibrators}.
\newblock {\it Publ. Astron. Soc. Aus.\/} {\bf 17}, 72 (2000).

\bibitem{Falcke95I}
H.~{Falcke}, P.~L. {Biermann}, {The jet-disk symbiosis. I. Radio to x-ray
  emission models for quasars.}
\newblock {\it Astron. Astrophys.\/} {\bf 293}, 665 (1995).

\bibitem{Herrnstein97}
J.~R. {Herrnstein}, {\it et~al.\/}, {Discovery of a subparsec jet 4000
  schwarzschild radii away from the central engine of NGC 4258}.
\newblock {\it Astrophys. J.\/} {\bf 475}, L17 (1997).

\bibitem{Kumar13}
P.~{Kumar}, R.~{Barniol Duran}, {\v Z}.~{Bo{\v s}njak}, T.~{Piran}, {A model
  for the multiwavelength radiation from tidal disruption event Swift
  J1644+57}.
\newblock {\it Mon. Not. R. Astron. Soc.\/} {\bf 434}, 3078 (2013).

\bibitem{Sokolovsky11}
K.~V. {Sokolovsky}, Y.~Y. {Kovalev}, A.~B. {Pushkarev}, A.~P. {Lobanov}, {A
  VLBA survey of the core shift effect in AGN jets. I. Evidence of dominating
  synchrotron opacity}.
\newblock {\it Astron. Astrophys.\/} {\bf 532}, A38 (2011).

\bibitem{Stone+13}
N.~{Stone}, R.~{Sari}, A.~{Loeb}, {Consequences of strong compression in tidal
  disruption events}.
\newblock {\it Mon. Not. R. Astron. Soc.\/} {\bf 435}, 1809 (2013).

\bibitem{Nakar&Piran11}
E.~{Nakar}, T.~{Piran}, {Detectable radio flares following gravitational waves
  from mergers of binary neutron stars}.
\newblock {\it Nature\/} {\bf 478}, 82 (2011).

\bibitem{Granot+02}
J.~{Granot}, A.~{Panaitescu}, P.~{Kumar}, S.~E. {Woosley}, {Off-axis afterglow
  emission from jetted gamma-ray bursts}.
\newblock {\it Astrophys. J.\/} {\bf 570}, L61 (2002).

\bibitem{VanEerten+10}
H.~{van Eerten}, W.~{Zhang}, A.~{MacFadyen}, {Off-axis gamma-ray burst
  afterglow modeling based on a two-dimensional axisymmetric hydrodynamics
  simulation}.
\newblock {\it Astrophys. J.\/} {\bf 722}, 235 (2010).

\bibitem{Wygoda+11}
N.~{Wygoda}, E.~{Waxman}, D.~A. {Frail}, {Relativistic jet dynamics and
  calorimetry of gamma-ray bursts}.
\newblock {\it Astrophys. J.\/} {\bf 738}, L23 (2011).

\bibitem{Panaitescu&Kumar01}
A.~{Panaitescu}, P.~{Kumar}, {Fundamental physical parameters of collimated
  gamma-ray burst afterglows}.
\newblock {\it Astrophys. J.\/} {\bf 560}, L49 (2001).

\bibitem{Metzger12}
B.~D. {Metzger}, D.~{Giannios}, P.~{Mimica}, {Afterglow model for the radio
  emission from the jetted tidal disruption candidate Swift J1644+57}.
\newblock {\it Mon. Not. R. Astron. Soc.\/} {\bf 420}, 3528 (2012).

\bibitem{BarniolDuran&Piran13}
R.~{Barniol Duran}, T.~{Piran}, {On the origin of the radio emission of Sw
  1644+57}.
\newblock {\it Astrophys. J.\/} {\bf 770}, 146 (2013).

\bibitem{Stone&Loeb12}
N.~{Stone}, A.~{Loeb}, {Observing Lense-Thirring precession in tidal disruption
  flares}.
\newblock {\it Physical Review Letters\/} {\bf 108}, 061302 (2012).

\bibitem{Bromberg+11}
O.~{Bromberg}, E.~{Nakar}, T.~{Piran}, R.~{Sari}, {The propagation of
  relativistic jets in external media}.
\newblock {\it Astrophys. J.\/} {\bf 740}, 100 (2011).

\end{thebibliography}
\end{document}